\RequirePackage{fix-cm}

\documentclass{svjour3}                    
\smartqed  

\usepackage{graphicx}
\usepackage{amsmath}
\usepackage{amssymb}
\usepackage[utf8]{inputenc}
\usepackage{natbib,amssymb}
\usepackage{pdfpages} 
\usepackage{comment}
\usepackage{textgreek}
\newcommand{\TL}[1]{\textcolor{black}{#1}}
\newcommand{\eb}[1]{\textcolor{black}{#1}}
\newcommand{\E}[1]{\textcolor{black}{#1}}
\journalname{Experiments in Fluids}
\begin{document}

\title{Space-time statistics of 2D soliton gas in shallow water studied by stereoscopic surface mapping}

\author{T. Leduque \and E. Barth\'elemy \and H. Michallet \and J. Sommeria \and N. Mordant}

\institute{Laboratoire des Ecoulements G\'eophysiques et Industriels, Universit\'e Grenoble Alpes, CNRS,
Grenoble-INP, F-38000 Grenoble, France
}

\date{Received: 26 September 2023 / Accepted: 8 May 2024}

\maketitle

\begin{abstract}
We describe laboratory experiments in a 2D wave tank that aim at building up and monitor 2D  shallow water soliton gas. The water surface elevation is obtained over a large ($\sim 100\,\text{m}^2$) domain, with centimetre-resolution, by stereoscopic vision using two cameras. Floating particles are seeded to get surface texture and determine the wave field by image correlation. 
\TL{
With this set-up, soliton propagation and multiple interactions can be measured with a previously 
unreachable level of detail.
}
The propagation of an oblique soliton is analysed, the amplitude decay and local incidence are compared to analytical predictions. We further present two cases of 2D soliton gas, emerging from multiple  line solitons with random incidence ($|\theta|<30^\circ$) and from irregular random waves forced with a {\sc jonswap} spectrum ($|\theta|<45^\circ$). 
\eb{To our knowledge, those are the first observations of random 2D soliton gas for gravity waves.} 
In both cases Mach reflections and Mach expansions result in solitons that mainly propagate in directions perpendicular to the wave-makers. 
\end{abstract}

\section*{Graphical abstract}
\hspace*{2cm}{\includegraphics[width=0.7\linewidth,viewport=0 0 1000 600,clip]{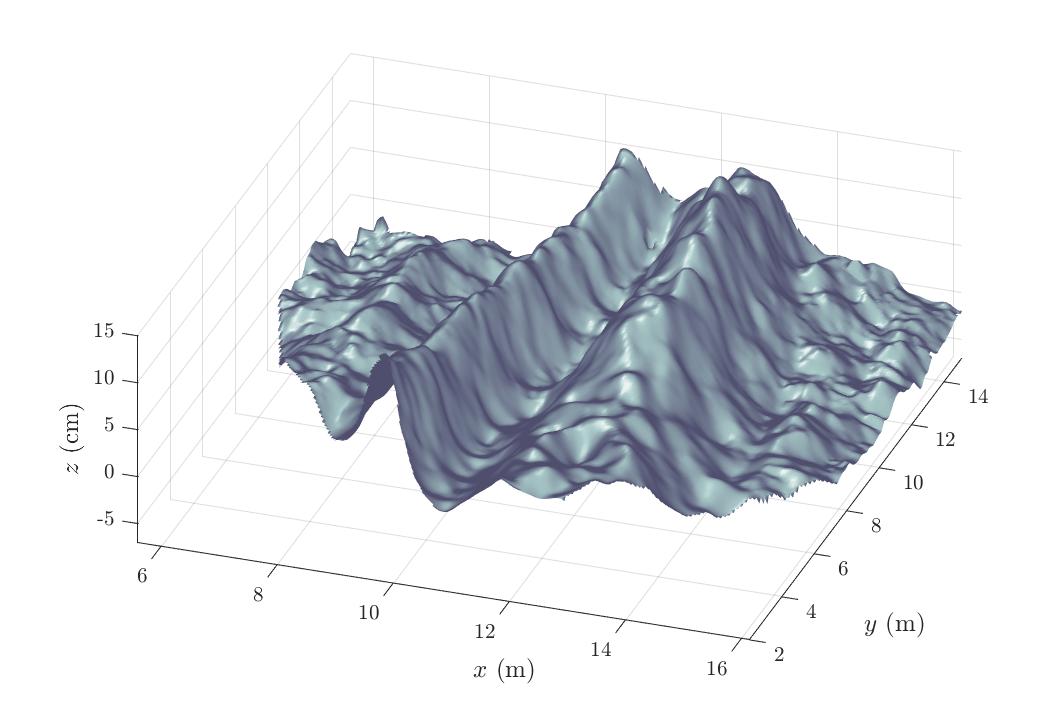}}

\section{Introduction}

\cite{zabusky1965interaction} used for the first time the word ``soliton" to characterize pulses that reappear virtually unaffected in shape and size after interaction. Solitons are solutions of integrable nonlinear dispersive equations, such as the Korteweg-de Vries (KdV), nonlinear Schr\"odinger, Benjamin-Ono, and many other equations. These equations find applications in different fields of physics, such as nonlinear optics, water surface waves, plasma waves, condensed matter \cite[]{dauxois2006physics}. 
\TL{
 The statistical properties of ensemble of solitons, randomly distributed in amplitude and position, were studied theoretically since then \cite[e.g.]{zakharov1971,el2005kinetic, congy2021soliton} and led to numerous fundamental studies \cite[e.g.]{bonnemain2022generalized, el2020spectral, gelash2019bound}. 
}
\cite{zakharov1971} named such an ensemble of solitons, ``soliton gas'', by analogy with particles colliding in a gas.

\cite{kadomtsev1970stability} derived a two-dimensional extension of the KdV equation to study soliton propagation with weak angular spreading. The Kadomtsev–Petviashvili (KP) equation is an integrable equation and admits exact solutions, including localized solitons along distinct lines in the horizontal plane (``line solitons''), and interactions between multiple line solitons that form two-dimensional patterns \cite[]{kodama2016kp}. Most interesting features arise for small incidence angles. The interaction with a vertical wall with an acute angle ($<30^\circ$) is known to produce Mach reflection as described theoretically by \cite{miles1977obliquely} and confirmed experimentally \cite[]{melville1980,li2011}. On theoretical grounds obtuse incidence ($>90^\circ$) leads to soliton diffraction or Mach expansion also a feature of the evolution of a finite soliton crest \cite[]{ryskamp2021} but not confirmed experimentally so far.

To produce soliton gases experimentally, several issues have to be faced. The main one is wave energy dissipation as in any experimental set-up. A stationary regime can only be reached by injecting energy with a forcing device. These two features, dissipation and forcing, are at odds with the soliton gas framework. Only recently \cite[]{redor2019experimental,redor2020analysis,redor2021} showed evidence of soliton gas forming in a shallow water wave flume \eb{(1D)} experiment. 
In their set-up, dissipation is acting on a much larger time scale compared to that of dispersion and non-linearity:  solitons emerge from monochromatic forcing and their dynamics can be considered adiabatic. 
\TL{Unidirectional propagating soliton gases have been generated in optic fibers \cite[]{marcucci2019topological, suret2023soliton}, or in deep water \cite[]{suret2020nonlinear}. In these studies initial conditions, computed in the framework of the non-linear Schr\"odinger equation, experimentally evolve into envelop solitons giving rise to localized dense soliton gases.}

In this article, we describe a large scale experiment that is designed to build up and monitor 2D  shallow water soliton gas in a ($30 \times 27$\,m$^2$) laboratory wave tank. \eb{The size of the investigation area and the time resolution are chosen for studying the dynamics of pure gravity waves (capillarity being negligible at these scales) in all horizontal directions and covering large ranges of wave numbers and frequencies.} For that purpose, the free surface elevation is video recorded over a large ($\sim 100\,\text{m}^2$) domain
, at high-rate (25\,Hz) during long times (several tens of minutes). Among all the techniques that were developed to measure the water free surface displacements \cite[for a review see][]{gomit2022free}, we apply a stereoscopic vision method developed by \cite{aubourg2019}. Two calibrated cameras record the same area. Image correlation is used to reconstruct the 3D wave field. Image correlation requires surface texture, for instance \cite{benetazzo2006measurements}, \cite{mironov2012statistical} and \cite{bergamasco2017wass} used the natural texture of the ocean made of short wind waves, capillary ripples and foam. For long non-breaking waves such as solitons, the water surface is very smooth. To circumvent this feature the tank was seeded with floating particles, following many other studies \cite[see e.g.][]{douxchamps2005,chatellier2013parametric, ferreira2017,aubourg2019}. 

The experimental set-up is described in the following section, along with the validation of the stereoscopic system against wave gauge measurements. In section 3, the case of a truncated oblique soliton is analysed, the amplitude decay is compared to the Mach expansion analytical solution of \cite{ryskamp2021}. We further present two cases of 2D soliton gas, emerging from a multiple line soliton generation and from irregular random waves forced with a {\sc {\sc jonswap}} spectrum.

\section{Experimental setup}
    \subsection{Wave tank}
		
        \begin{figure}
            \centering
            {\bf (a)}
            \hspace{5cm}
            {\bf (b)}
            \includegraphics[width=0.9\linewidth]{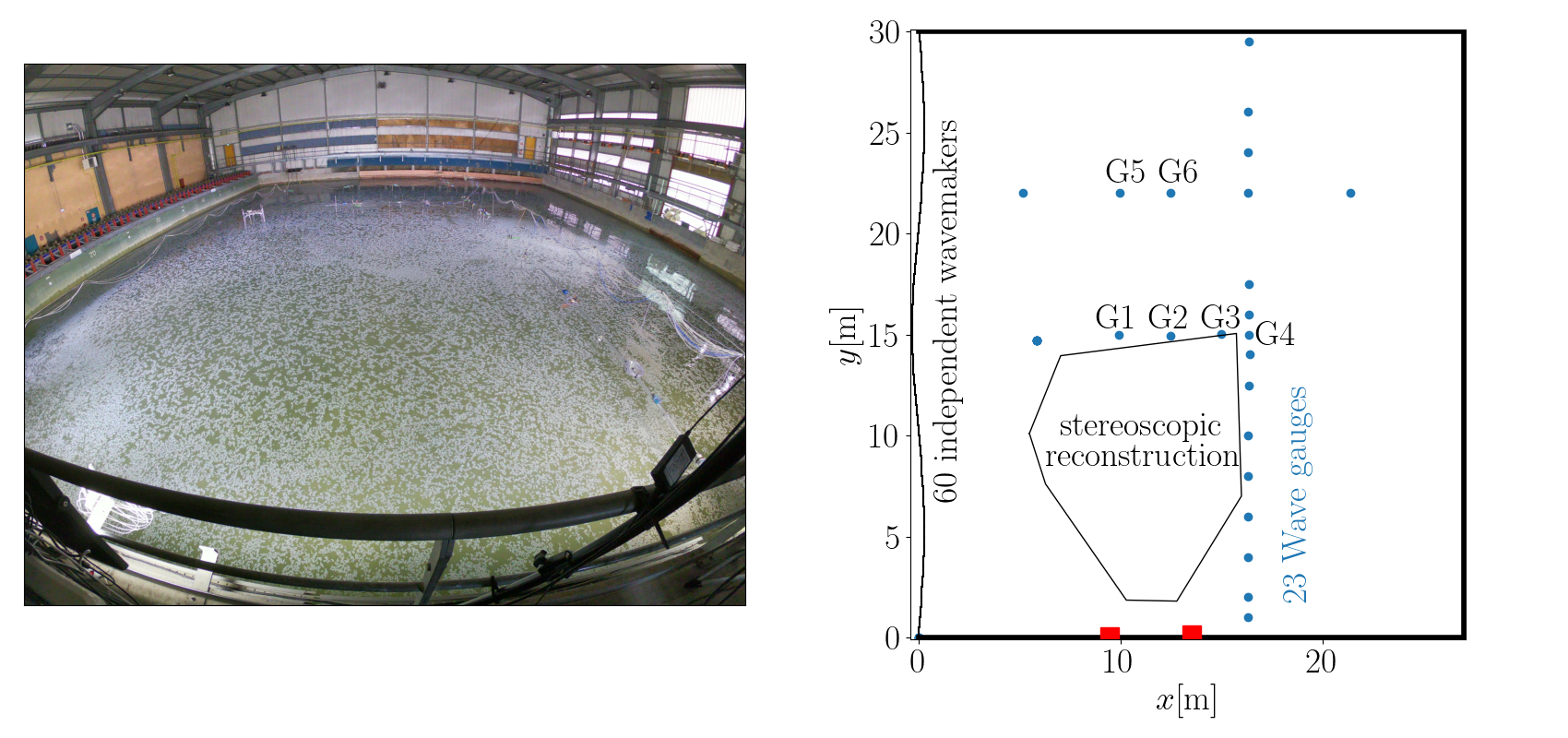}
            \caption{\textbf{a} View of the wave tank. The wave-makers are located on the left side of the picture. The water surface is covered with buoyant plastic particles to generate patterns for image correlation. \textbf{b} Top view sketch of the wave tank. The wall at $x=0$ is made of $60$ independent piston-type wave-makers. The water surface elevation is measured using $23$ wave gauges (blue points), 6 of them are numbered for reference in the text  (their coordinates are given in Table~\ref{tab:Gauge_Coord}). The elevation is also measured by a stereoscopic surface mapping technique over an area roughly 100~m$^2$ shown in the figure. This measurement is done using two cameras which positions are sketched by the two red rectangles. The cameras are 4.5 m above the tank bottom. 
            }
            \label{fig:WaveTank}
        \end{figure}

        The experiments were performed in the LHF wave tank located in Pont-de-Claix (France) which is operated by the ARTELIA company (Fig.~\ref{fig:WaveTank}). The tank is $30$~m wide and $27$~m long (between the mean position of the pistons and the vertical end wall). The water depth is set to $h=35$~cm for the experiments presented herein. The bottom is horizontal 
        \eb{with a standard deviation of $4$\,mm}
        over the entire surface. Three sides ($y=0$, $x=27\,$m and $y=30\,$m) of the tank are vertical walls so that the reflection coefficient of the waves is $1$. The wall at $x=0$ consists in sixty $50$~cm-wide pistons, with a maximum stroke of $0.6\,$m. The pistons are controllable in position independently of each other. Arbitrary multidimensional waves can be generated such as a single oblique soliton, multiple oblique solitons, oblique sine waves or multi-directional random forcing. This paper will focus on two types of wave forcing: solitons (single or multiple) and random waves using a {\sc jonswap} forcing 
        \TL{\cite[]{hasselmann1973measurements}}.

    \subsection{Wave generation} \label{Forcing}
    
    \subsubsection{Solitons}
        In the case of a 1D soliton, \cite{Guizien2002} provide the law of motion of a piston wave-maker for a Rayleigh type soliton \TL{also solution of the Serre-Green-Naghdi equations \cite[]{serre1953contribution}}. It describes a fully non-linear solitary wave that is recognized to shed less dispersive waves during propagation. An oblique Rayleigh soliton with an angle of propagation $\theta$ with respect to the $y$ axis has the following expression:
        
        \begin{gather} 
            \eta(x,y,t)=a_0 \, \text{sech}^2 \left( \beta (x\cos\theta + y\sin\theta -c t) \right)
            \label{eq:sol}
            \\
            \beta = \sqrt{\dfrac{3\, a_0}{4\, h^2(a_0+h)}} \\
            c = \sqrt{g\, (h + a_0)} 
		      \label{eq:sol_c}
        \end{gather}
        where $a_0$ is the amplitude of the soliton, $\beta$ the shape factor and $c$ is the phase speed. Following the same method as in \cite{Guizien2002}, the displacement $X(t)$ of each piston located at $y=y_i$ is computed with 
        
        \begin{equation}
            \frac{dX}{dt}(y_i,t)= c \, \frac{\eta\left(X(y_i,t),y_i,t\right) }{h+\eta\left(X(y_i,t),y_i,t\right)}\cos\theta \,.
			\label{eq:sol_gen}
        \end{equation}
        This differential equation is integrated numerically to obtain the piston displacement.
        
        \eb{The generation of a single soliton requires a net forward push of the wave-makers.}
        Multiple soliton generation is obtained by sequentially 
        \eb{emitting}
        single solitons 
        \eb{while the wave-makers continuously recede slowly at all times.}
        \eb{Preliminary tests carried out in a 1D wave flume show that the receding velocity of the piston is an important parameter} \eb{\cite[]{Leduque2021}}. It tunes the density of solitons \TL{(understood as the number of solitons emitted per unit time)}, the amount of spurious waves, and potential soliton amplification (when a new push matches the arrival time of a reflected soliton) \eb{which can cause wave breaking}. We typically use a receding velocity between 1 and 2~cm/s. \eb{Above 2~cm/s, soliton amplification leads to excessive wave breaking.}
        \eb{The instant of emission during the receding phase,}
        the angle $\theta$ and/or the amplitude $a_0$ 
        can be changed from one soliton to another in order to introduce some variability in the forcing.
        
        \begin{table}
            \begin{tabular}{c|c|c|c|c|c|c}
                & G1 & G2 & G3 & G4 & G5 & G6 \\
                \hline
                $x$ [m] & 9.92  & 12.46 & 15.01 & 16.36 & 9.95  & 12.49\\
                $y$ [m] & 14.99 & 14.95 & 15.01 & 15.00 & 22.00 & 22.01
            \end{tabular}
            \caption{Coordinates of wave gauges G1 to G6 sketched in Fig.~\ref{fig:WaveTank}b.
            }
            \label{tab:Gauge_Coord}
        \end{table}

    \subsubsection{Random forcing} \label{sect:generation:random}
    
    The {\sc awasys} software that drives the wave-maker has a builtin generation of multi-directional random waves complying with a {\sc jonswap} spectrum with angular spreading. The {\sc jonswap} spectrum has been developed to model the sea state in deep water \cite[see][]{hasselmann1973measurements}. We use a common $\cos^{2s}$ angular spreading distribution  \cite[see][]{goda1999comparative}. In the shallow water case, the relevance of the {\sc jonswap} spectrum is disputable but it is a convenient tool to generate multi-directional random waves. \TL{The {\sc jonswap} spectrum model and parameters are described in Appendix.}
    The forcing spectrum is strongly peaked at a frequency $f_p$ and is relatively narrow (with a $1/f^5$ decay at high frequency). The directivity of the spectrum can be tuned from unidirectional to $\pm 45^\circ$. Table~\ref{tab:Exp_para} summarizes the parameters of the {\sc jonswap} experiments A and B that are presented here \TL{(the forcing spectra will be compared to those obtain from the measurements in the next section). 
    } Experiments A and B are selected amongst $31$ different sets of tested {\sc jonswap} conditions because they resulted in random soliton gases most appropriate to validate the experimental procedures. A broader study of the different statistical regimes is planned in a forthcoming experimental campaign.

    \begin{table}
        \begin{tabular}{c|cccc}
            Experiment & $f_p$ & $H_{m0}$  & $\gamma$ & $\sigma_\theta$  \\
                        & [Hz] & [cm] &  & $[^\circ]$ \\
            \hline
                        A & 0.2 & 5.2       & 3.3 & $\pm 14$ \\
                        B & 0.2 & 3.5       & 10  & $\pm 25$ \\
        \end{tabular}
        \caption{Parameters of 
        {\sc jonswap} experiments: $f_p$ is the peak frequency of the spectrum of the forced waves, $H_{m0}$ the significant wave height of the forcing spectrum, $\gamma$ is a peak enhancement parameter and $\sigma_\theta$ is the standard deviation of the angular spreading around $\theta=0$ \cite[]{awasys}. 
        }
        \label{tab:Exp_para}
    \end{table}

\begin{figure}
        \centering
        {\includegraphics[width=0.8\linewidth,viewport=60 680 1250 1250,clip]{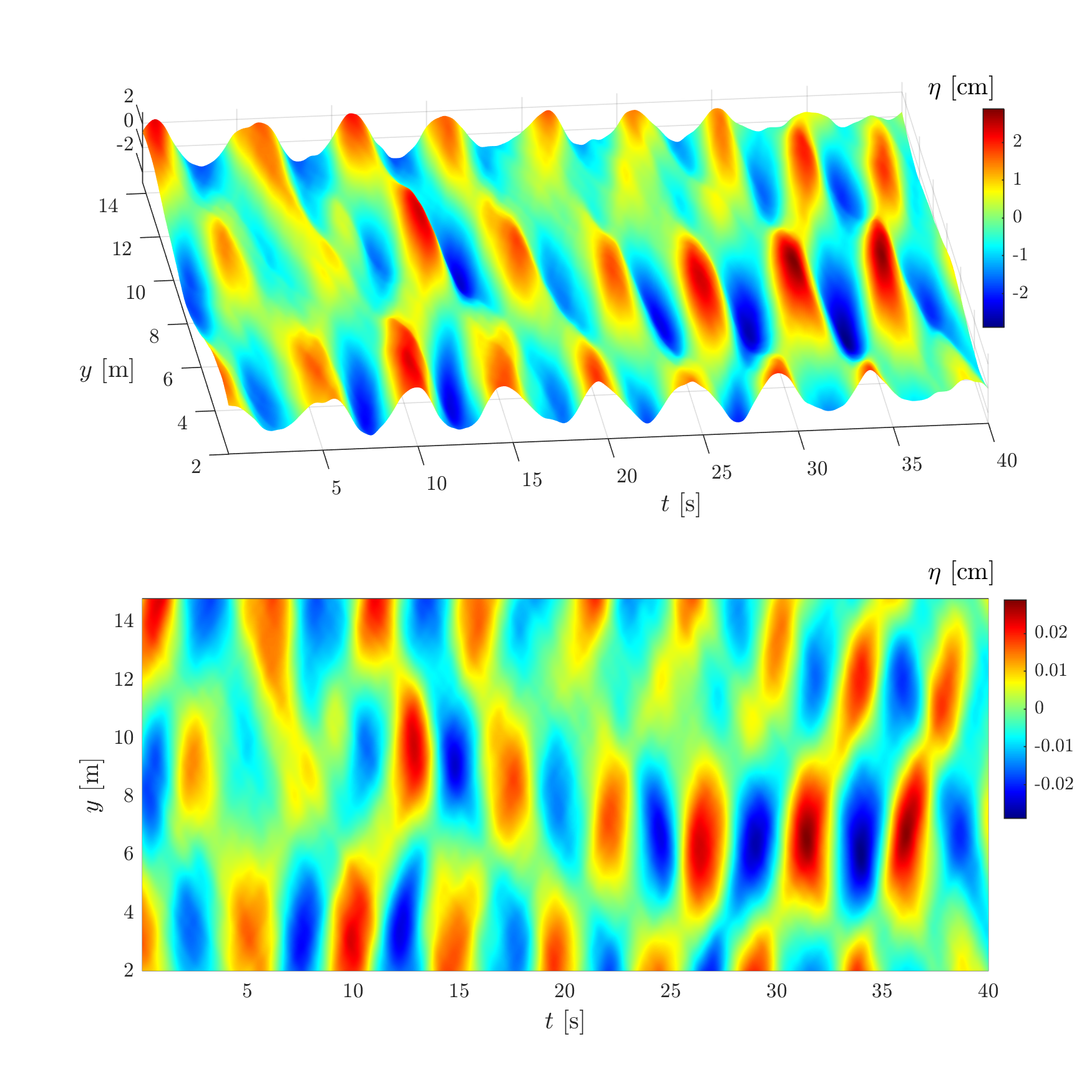}}
        \caption{Example of a portion of the surface elevation field $\eta(t,y)$ at the wavemakers ($x=0$) of the {\sc jonswap} forcing for experiment B.}
        \label{fig:jonswapforcing}
    \end{figure}

    \E{An example of the prescribed free surface elevations at the wavemaker is shown in Fig.~\ref{fig:jonswapforcing}. This space-time free surface elevation field would be the one generated in deep water in absence of any reflected waves. This theoretical field is the sum of many sinusoidal components, therefore the distribution of elevations is exactly a normal distribution. This example illustrates in particular the forcing for experiment B with a 5\,s peak period and a large directional spreading. The actual observed field in the tank is different due to the continuous forcing of the wave makers, reflections on the walls and non-linear interactions and will be discussed later in section~\ref{RandVal}.}

    \subsection{Water elevation measurements}

     Water surface displacements are recorded at fixed points by wave gauges and over an extended surface by stereoscopic surface reconstruction.

    \subsubsection{Wave gauges}
        
        Up to $23$ capacitive wave gauges are used (see blue points in Fig.~\ref{fig:WaveTank}b). They are easily deployed and provide an excellent accuracy of better than
\eb{1 mm.}
        Their calibration is very stable in time,
\eb{the precision after several days of experiments remains better than 2\,\%}. 
        The positions of gauges numbered from G1 to G6 are given in table \ref{tab:Gauge_Coord}. 
        \begin{figure}
            \centering
            {\bf (a)}            
            \includegraphics[width=10cm]{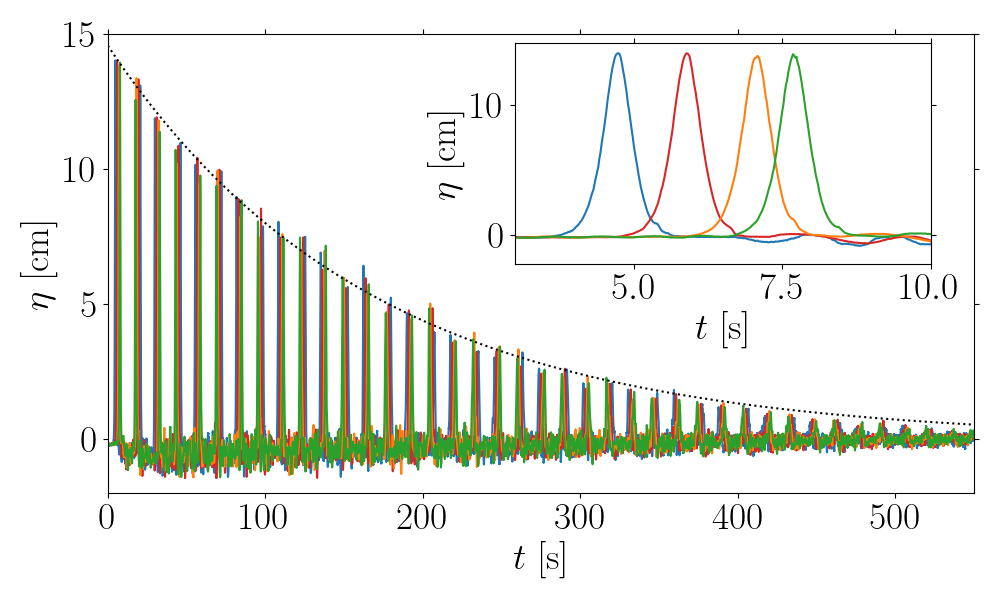}
              
            \hspace{.7cm}
            {\bf (b)}            
            \includegraphics[width=10cm]{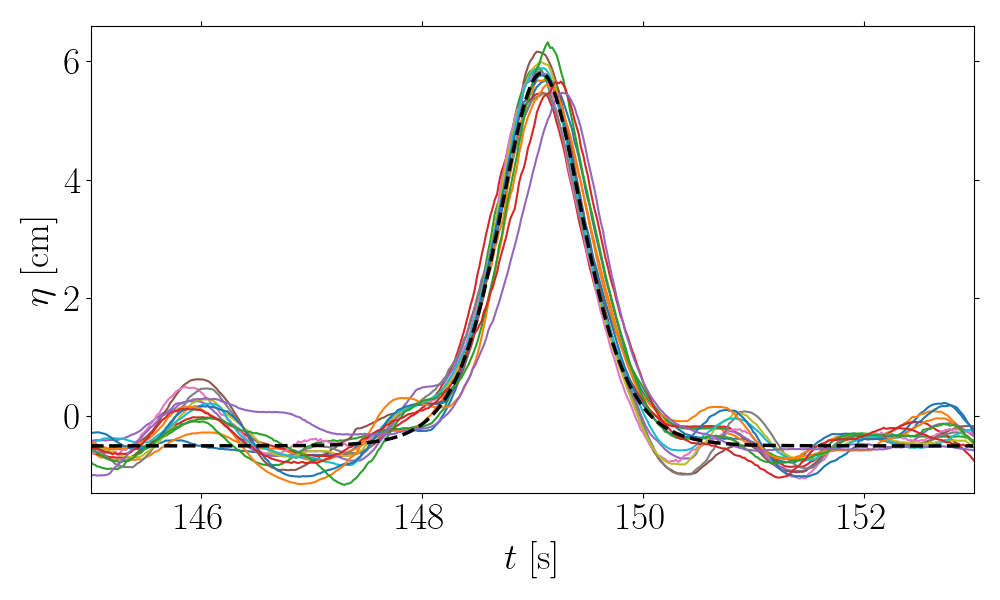}
            \caption{Propagation of a soliton of dimensionless amplitude ${a_0}/{h}=0.4$ with $\theta=0$. \textbf{a} Water surface elevations recorded with four wave gauges (see Fig.~\ref{fig:WaveTank}b): G1 in blue, G2 in red, G3 in orange, G4 in green. The insert is a focus on the beginning of the run before the first reflection at  $x=27\,$m. The dotted black line is the exponential fit of the soliton decay by relation (\ref{exp:fit}) with $\tau=167\,$s. \textbf{b} Water surface elevations recorded with 15 wave gauges, all located at $x= 16.3\,$m. The black dashed line is the Rayleigh soliton of amplitude $a=6.3$~cm.
            }
            \label{fig:sol_keulegan}
        \end{figure}
        
        Fig.~\ref{fig:sol_keulegan}a illustrates the propagation of a single soliton produced by all pistons moving in phase. This configuration is similar to the case of 1D propagation in a narrow wave flume as the wave crest is parallel to the wave-maker wall. The water surface displacements are shown at four locations (gauges G1, G2, G3 and G4). The soliton propagates back and forth with reflection on the $x=27\,$m wall and on the wave-maker. Energy dissipation by viscous friction on side walls and bottom is the main mechanism of amplitude attenuation at long times \citep{keulegan1948}. In such a large tank, the friction on the side walls during propagation is negligible. Additional attenuation occurs as large solitons interacting with the end-walls produce small dispersive waves, similarly to head-on collision of two solitons \citep[see e.g.][]{Chen2014}. The amplitude $a$ empirically decreases according to an exponential law \citep{redor2019thesis}
        \begin{equation}
            a(t) = a_0 \, \exp{\left(-\frac{t-t_0}{\tau}\right)}  \,,
			\label{exp:fit} 
        \end{equation}
        with $a_0$ the soliton amplitude at time $t_0$ and $\tau$ the dissipation time scale. The best fit for this viscous dissipation time scale is $\tau_t = 167\,$s. The equivalent viscous dissipation length scale is $\sqrt{gh} \tau = 310\,$m. Before being undetectable due to its decay, the soliton propagates over $34$ tank lengths. At each passage, the soliton crest elevation varies by less than a few percents at a same transverse location, see Fig.\ref{fig:sol_keulegan}b (all along the line of gauges at $x=16.34\,$m sketched in Fig.~\ref{fig:WaveTank}). These fluctuations originate most likely from the slight tank bottom 
        \eb{unevenness that may trigger weak transverse  amplitude modulations.}
        This shows nonetheless the global stability of such a single line soliton. 

    \subsubsection{Stereo-video}
        \eb{Stereoscopic reconstruction of the water surface elevation is based on image cross-correlation between the simultaneous video frames of two calibrated cameras. We use two synchronized s{\sc cmos} cameras (pco.edge 5.5) running at 25 frames per second with $5.5\,$Mpixels images. 
        }
        Various constraints arise for such a spatial measurement: wave amplitudes range from few millimetres high linear waves up to $25\,$cm for nearly breaking waves, horizontal length scales range from $10\,$cm for linear waves ($\sim 1\,$m for a soliton) to tank-scale waves. This calls for a large measurement area with a spatial horizontal resolution of at least $10\,$cm and an elevation resolution better than $1\,$cm. To meet such requirements $14\,$mm camera lens were chosen. The calibration is an important step to correct the significant distortion induced by the short focal length lenses and the large distances. The calibration is done using the {\sc Matlab} Toolbox described in \cite{zhang2000}. It results in $12$ parameters (for one camera). Among them 6 intrinsic parameters are used for the optical system that is approximated by the ideal pinhole camera model with fourth order radial distortion. 
 \eb{For each cameras these parameters are obtained with ten images of a grid test pattern made of 700 points.}
 The other 6 parameters are extrinsic. 
 They express the three translations and the three angles of the rotation matrix of the cameras with respect to the wave tank coordinate system.
 \eb{The extrinsic parameters are determined using 35 points with known coordinates distributed in the common field of view of the two cameras.}
        
        \TL{The principle of stereo reconstruction relies on the recognition of a same surface pattern viewed from two lines of sight, by image correlation. Even in the presence of waves the water surface in the experiments is not rough enough at small scale to make image correlation work. To obtain good correlation level \cite[following][e.g.]{douxchamps2005,chatellier2013parametric,ferreira2017}, we seed the water surface with buoyant particles (the white spots in Fig.~\ref{fig:WaveTank}a).} 
        We use {\sc peld} nearly spherical particles with diameter $3\,$mm and density $0.935$ kg/m$^3$.
        \eb{Preliminary agitation with short waves was used at the start of the experiments, in order to disrupt large holes and clusters, so to ensure the particles dispersion over the whole tank.
        The surface is illuminated with 15 high-power spot-lights (70\,kW in
total) installed behind the cameras to avoid specular reflection issues.
        }

        \begin{figure}
            \centering
            \textbf{(a)}\includegraphics[width=7 cm]{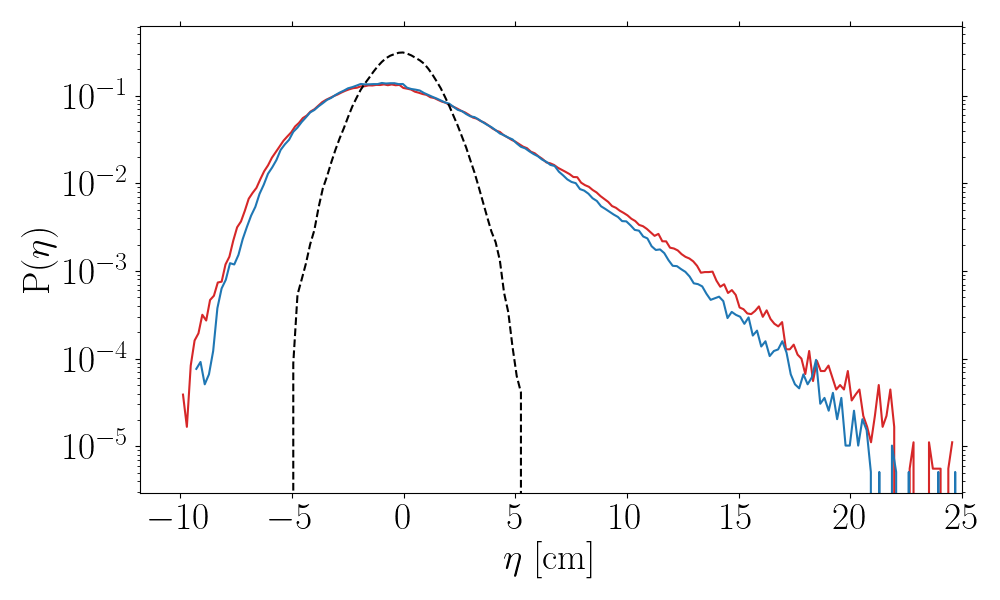}
            \hspace{5mm}
            \textbf{(b)}\includegraphics[width=7 cm]{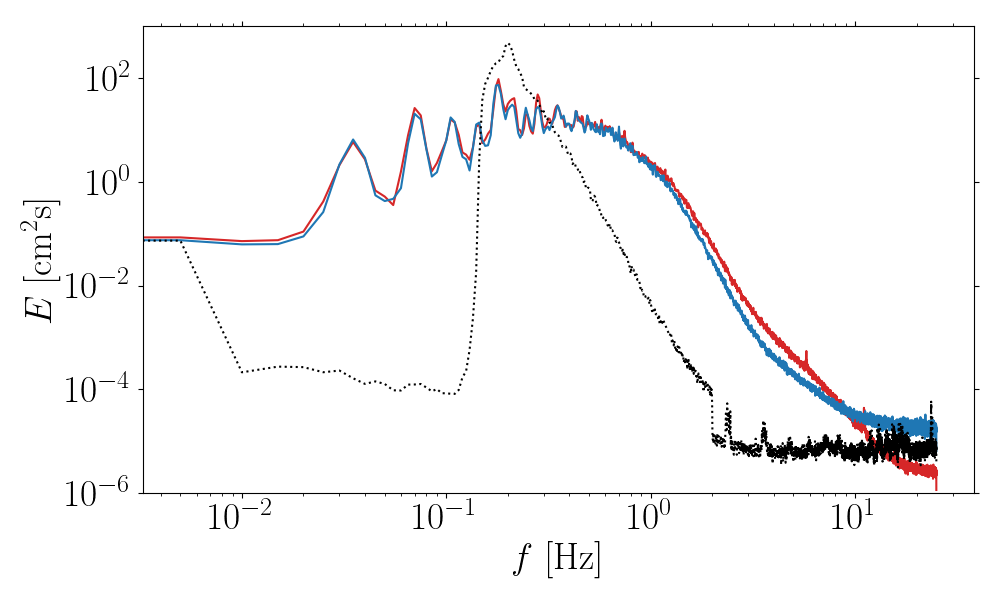}
            \caption{\textbf{a} Probability density function of the free surface elevation \eb{measured by 13 wave gauges} and \textbf{b} Fourier power density spectra for {\sc jonswap} experiment A, carried out with or without particles (blue and red curves, respectively). 
            \eb{The dashed black line in \textbf{a} is the probability density function of the surface elevation forced by the wavemaker at $x=0$. The dotted black line in \textbf{b} is the spectrum of the motion of the wave-maker.}
            }
            \label{fig:Comp_part}
        \end{figure}

        In order to check if the dynamics of the surface motions are not altered by the particles, experiment A (with {\sc jonswap} forcing) is run twice: with and without particles. In Fig.~\ref{fig:Comp_part}, we show the probability density function ({\sc pdf}) and Fourier power spectral density of the surface displacements. The {\sc pdf} curves are very close in the range of elevations between $-7\,$cm and $7\,$cm. Large wave crests seem to be slightly damped by the presence of the particles. Concerning the spectra, both curves closely match up to $f=1$ Hz. The particles do not modify the dynamics at low frequency $f<1\,$Hz but energy at higher frequencies is reduced as it would be with surfactants \cite[]{Campagne}. This is consistent with the amplitude attenuation described for the {\sc pdf} plots. With particles, the spectrum saturates at noise level for $f \sim 7$ Hz. Noteworthy is the power spectral density spanning 6 orders of magnitude. We thus conclude from this test that the addition of particles has marginal effects on the dynamics of the wave motions we intend to measure.

        The PDF distribution displayed in Fig.~\ref{fig:Comp_part}a is strongly skewed and the spectrum decay at high frequency in Fig.~\ref{fig:Comp_part}b is close to exponential. These two features are similar to that of shallow water 1D soliton gas \cite[][]{redor2021}. 2D soliton gas characteristics will be further discussed in section 3 based on the video data.

\begin{figure}
            \centering
            \includegraphics[width=.9\linewidth]{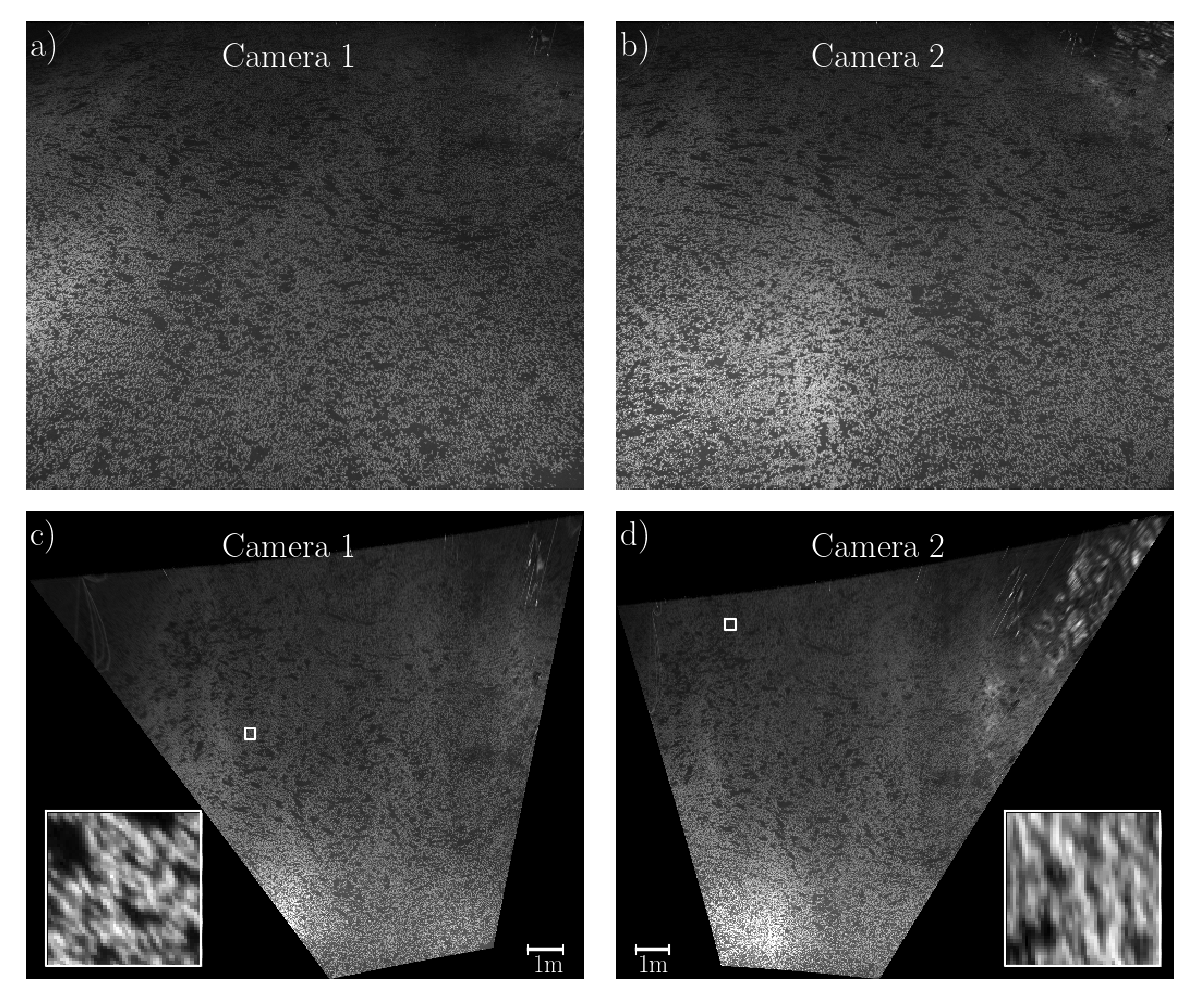}
            \caption{\textbf{a-b}: Pair of simultaneous camera images. \textbf{c-d}: \E{images \textbf{a-b} rectified: projected on the  plane $z=0$}. The inserts show enlargements of the $35 \times 35\,$cm$^2$ areas bounded by the white squares.
            }
            \label{fig:ImgCam}
        \end{figure}

        The stereoscopic surface mapping is undertaken using an in-house {\sc Matlab} toolbox called {\sc uvmat} described in \cite{aubourg2019} and in more detail in \cite{aubourg2016etude}. The first step is the rectification of the images \E{that is the projection on the $z=0$ plane}: examples are shown in Fig.~\ref{fig:ImgCam}. \eb{After rectification the images of the two cameras are identical if there is no vertical displacement. The purpose of this step is to correct image differences due to camera distortion and orientation \cite[see Appendix 5 of][]{aubourg2019}. In the rectified images (Figs.\,\ref{fig:ImgCam}c-d), the pixel size is 7~mm.} A multigrid iterative processing \cite[]{scarano1999iterative} is then performed. Each rectified image is split into 84~cm 
        square windows in which 31~cm 
        square boxes are searched for best correlation. The resulting virtual horizontal displacements constitute the input for a new interrogation based on a finer grid, where the search window and the correlation box are refined to 35~cm 
        and 
        21~cm, respectively. 
        \eb{The size of the last correlation box defines the smallest detectable wavelength, i.e. 42~cm}. The virtual horizontal displacements are then transformed into the elevation field \cite[see Eqs.~5-6 in][]{aubourg2019}. Locations where the correlation is poor are discarded. Poor correlation is the result of unclear patterns, that can be due to too dense or too low particle concentration locally, as a result of wave breaking for instance. We have to emphasize that the wave conditions were chosen to minimize wave breaking occurrence. The validated elevations are interpolated over a regular grid with a mesh size of $5 \times 5 \, \text{cm}^2$ using thin-plate spline functions. A mask is applied in order to restrict the measurement region because parts of the periphery of the reconstruction area are altered by reflections of light or by the presence of cables or gauges (see for example the top left and right part of Fig.~\ref{fig:ImgCam}c,d). 
        
        The following section focuses on the validation of this measurement technique by comparison with the wave gauge measurements.      
        
    \subsection{Validation of the stereoscopic surface reconstruction}
        
        \subsubsection{Case of random forcing} \label{RandVal}

        \begin{figure}
            \centering
            {\includegraphics[width=0.8\linewidth,viewport=60 680 1250 1250,clip]{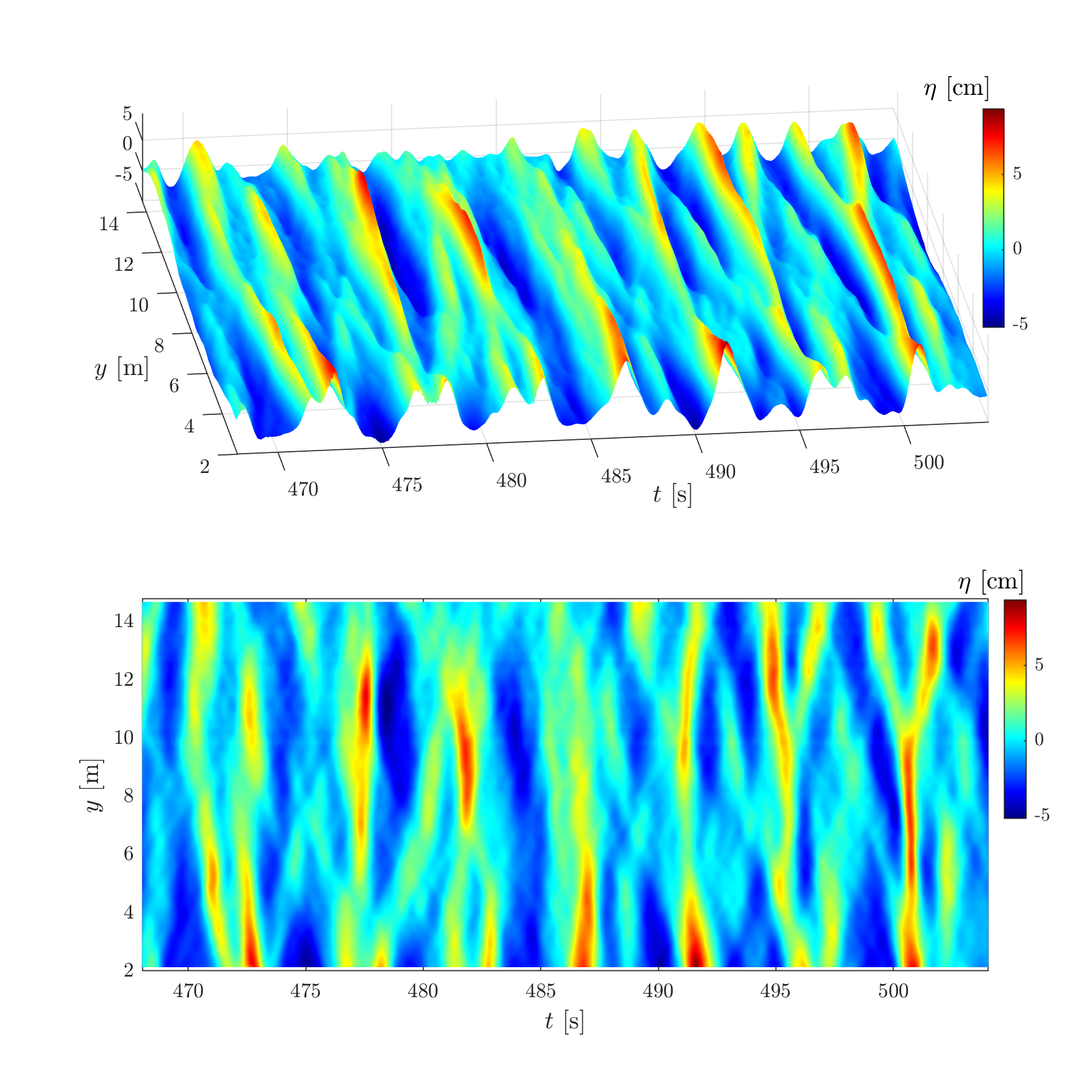}}
            \caption{A portion of the surface elevation field $\eta(t,y)$ at $x=13$\,m measured during experiment B.}
            \label{fig:etayt1199}
        \end{figure}
        
        \begin{figure}
            \centering
            {\bf (a)}\includegraphics[width=7cm]{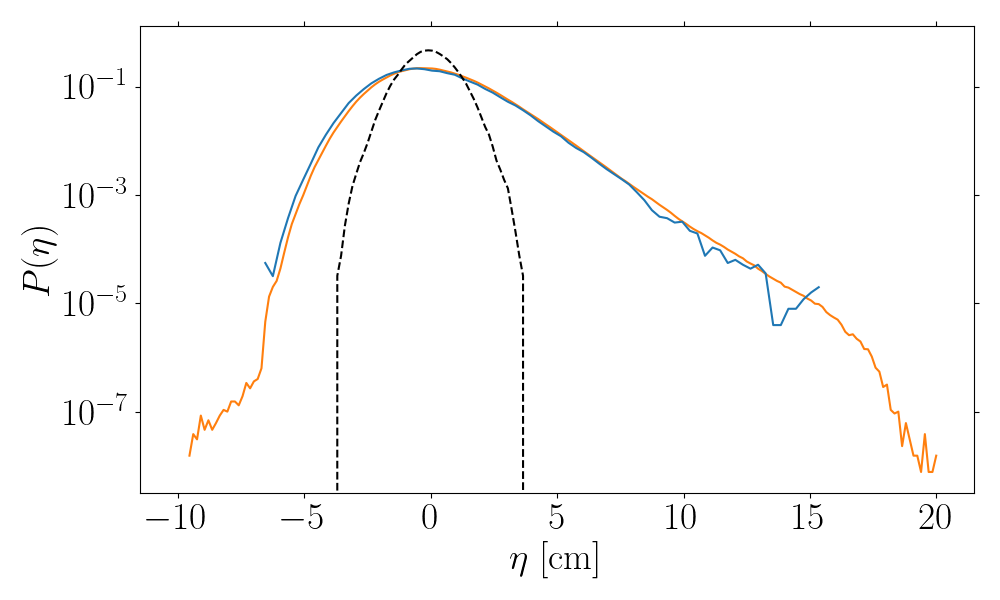}
            \hspace{5mm}
            {\bf (b)}\includegraphics[width=7cm]{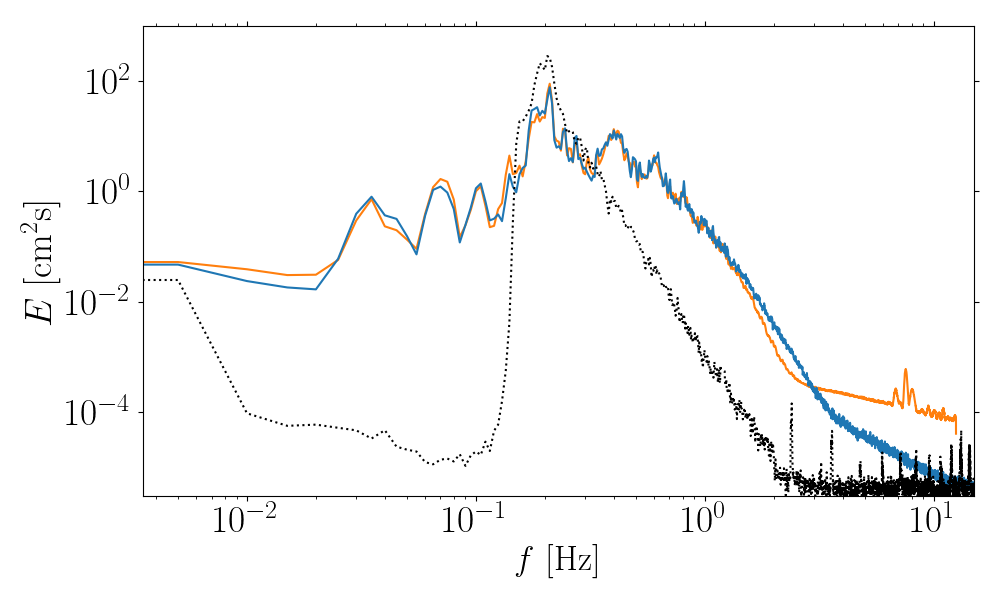}
            \caption{
				\textbf{a} Probability density function of the free surface displacements and \textbf{b} Fourier power density spectrum of the free surface displacements
                for  {\sc jonswap} experiment B, from the stereoscopic system (orange curves) and wave gauges (blue curves).
                \eb{The dashed black line in \textbf{a} is the 
    normal distribution
                of the surface elevation forced by the wave-maker at $x=0$ (see Fig.\ref{fig:jonswapforcing}).
                The dotted black line in \textbf{b} is the spectrum of the wave-maker displacement.}
                }
            \label{fig:val_stereo}
        \end{figure}
        
        \TL{An example of a portion of the wave field measured during experiment B is shown in fig.~\ref{fig:etayt1199}. This wave field is qualitatively very different from that of the forcing in Fig.~\ref{fig:jonswapforcing}. Large and peaky waves that resemble solitons are detected. This feature will be discussed in section~\ref{RandDyn}. In the present section we primarily intend to validate the accuracy of the stereo-video measurement.}
        For a random case, visual observation indicates that the wave field is rather homogeneous a few metres away from the walls. It is thus reasonable to compare the spectral and statistical properties of the 3D reconstruction to that given by wave gauges that are located out of the field of view of the cameras. The {\sc pdf} and the Fourier power spectral density of the water elevation for the {\sc jonswap} experiment B are displayed in Fig.~\ref{fig:val_stereo}. The {\sc pdf} is computed with a record duration of 14 minutes and averaged either on the wave gauges (blue curves) or over the entire stereoscopic reconstruction area (orange curves). {\sc pdf} estimates are shown in Fig.~\ref{fig:val_stereo}a. \eb{As one point of the stereoscopic reconstruction} provides the same amount of data than one gauge, the amount of data for the stereoscopic measurement is far larger than that of gauges. The {\sc pdf} from the stereoscopic reconstruction therefore appears to be far more converged  compared to that of the wave gauges. There is an overlapping range of elevations for which the two curves are very close.

        The corresponding power spectra are shown in Fig.~\ref{fig:val_stereo}b. For $f<1.5$~Hz, the two curves are very close. The spectra exhibit modal behaviour with low frequency peaks. The lowest is at $f \simeq 0.033\,$Hz, which corresponds to the first seiching mode of the tank. The other peaks below $0.02$~Hz are higher order seiching modes. The two estimates match very well for the short wave frequencies ($0.02<f<1.5$~Hz). For $1.5<f<2$~Hz, the spectrum  from the cameras decays slightly faster than that of the gauges showing low pass filtering. The noise level is reached around $f=2$~Hz for the stereoscopic measurement, slightly higher than that of the wave gauges (reached at $f\simeq 3$~Hz). \E{Of note, this cut-off at 2~Hz is linked to the
spatial resolution through the wave phase velocity given by the Airy relation.}
        A dynamical range of $5$ orders of magnitude in power density is achieved nonetheless.

        This comparison indicates that the statistical and spectral estimates of a random wave field are perfectly captured by the stereoscopic surface mapping.

        \subsubsection{Single oblique soliton experiment}

        \begin{figure*}
            \includegraphics[width=\linewidth]{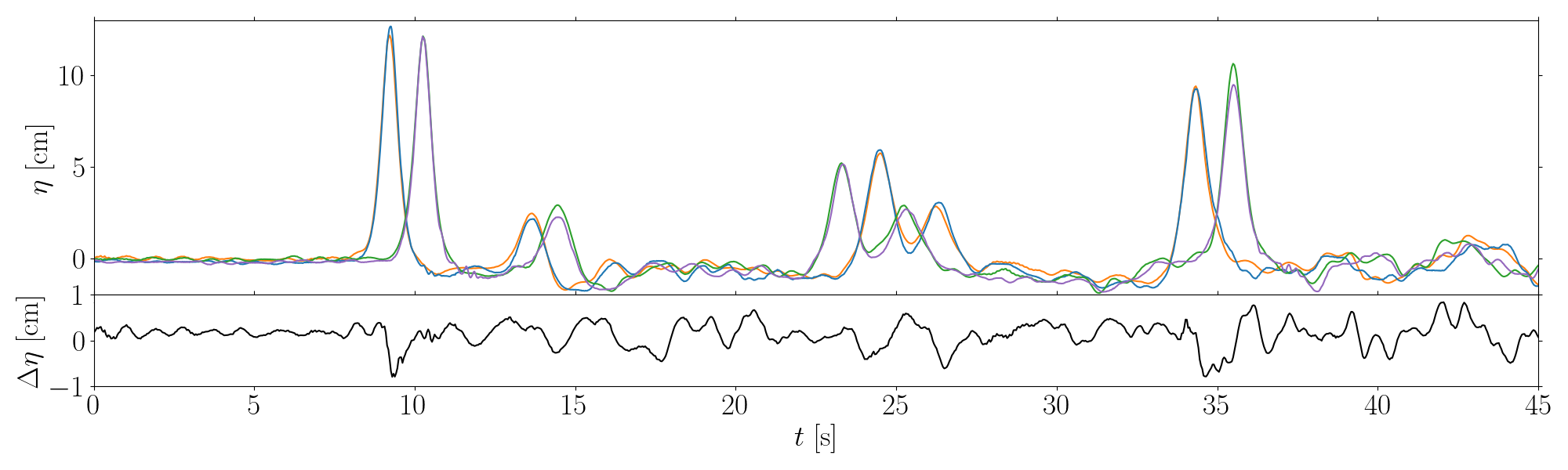}
            \\
            \vspace*{-4.8cm}
            \\
            \hspace*{1cm}
            \textbf{(a)}
            \vspace*{2.3cm}						
            \\
            \hspace*{1cm}
            \textbf{(b)}
            \vspace*{1cm}
            \caption{\textbf{a} Time series of water surface displacements for an oblique soliton, ${a_0}/{h}= 0.35$ and $\theta=\pm 30^\circ$. Wave gauges (blue and purple) at $y=22$\,m (run with $\theta=+30^\circ$) and virtual gauges from stereoscopic reconstruction (orange and green) at $y=8$\,m  (run with $\theta=-30^\circ$), at $x= 10$\,m (blue and orange) and $x= 12.4$\,m (purple and green). \textbf{b} Difference between stereoscopic and wave gauge measurements at $x= 10$\,m.
            }
            \label{fig:val_eta_2exp}
        \end{figure*}
        
        We aim at comparing stereoscopic mapping and wave gauge measurements for a single oblique soliton. Since there is no wave gauge in the stereoscopic measurement area (in order to clear the view of the cameras), a direct comparison of the two measurement systems is not possible. To overcome this issue we compare the measurements of two symmetrical runs, one with a single soliton inclined at $\theta = +30^\circ$ and the other one inclined at $\theta=-30^\circ$. In both runs, the soliton has the same reduced amplitude ${a_0}/{h}= 0.35$. Consequently, the time evolution of the first run in the half tank $0<y<15\,$m is  symmetric to that of the second run for the other half tank $15<y<30\,$m. Taking advantage of this symmetry, water surface elevations from the two systems are compared in Fig.~\ref{fig:val_eta_2exp}a. Both types of measurements are very similar. The difference $\Delta\eta$ between the elevations measured at $x=10$\,m is plotted in Fig.~\ref{fig:val_eta_2exp}b. Slight phase shifts in the wave records cause the main peaks in $\Delta\eta$. The root mean square (rms) difference between the two types of measurements is computed over 60 seconds. \TL{A rms value of $3$~mm is found that gives the order of magnitude of the vertical resolution of the video system.} The rms difference between the elevations measured at two wave gauges located symmetrically at $x=16.34$\,m for the two symmetric runs is of the same order of magnitude. It is not solely the estimation of measurement accuracy. It also encompasses the differences in symmetric paddle motions, inaccuracy in gauges location, and non-uniformity of the tank bottom. The latter induces slight variations in wave speeds \eb{(up to 2\,\% locally say, for a maximum bed variation of 1\,cm)}. All these uncertainties contribute to the slight differences in soliton arrival time on symmetric locations. 

        The comparison of the performance of the two measurement techniques shows that the stereoscopic reconstruction performs very well with a slight decay of the dynamical range due to higher noise level and a slight low pass filtering of the highest frequencies. The obvious advantage of the video measurement is to combine high resolution in both time and space. The added value of the spatial resolution is illustrated on three examples in the next section.

\section{Stereoscopic wave field data analysis}

    \subsection{Single oblique soliton} \label{1SolDyn}

       \begin{figure}
            \includegraphics[width=\linewidth]{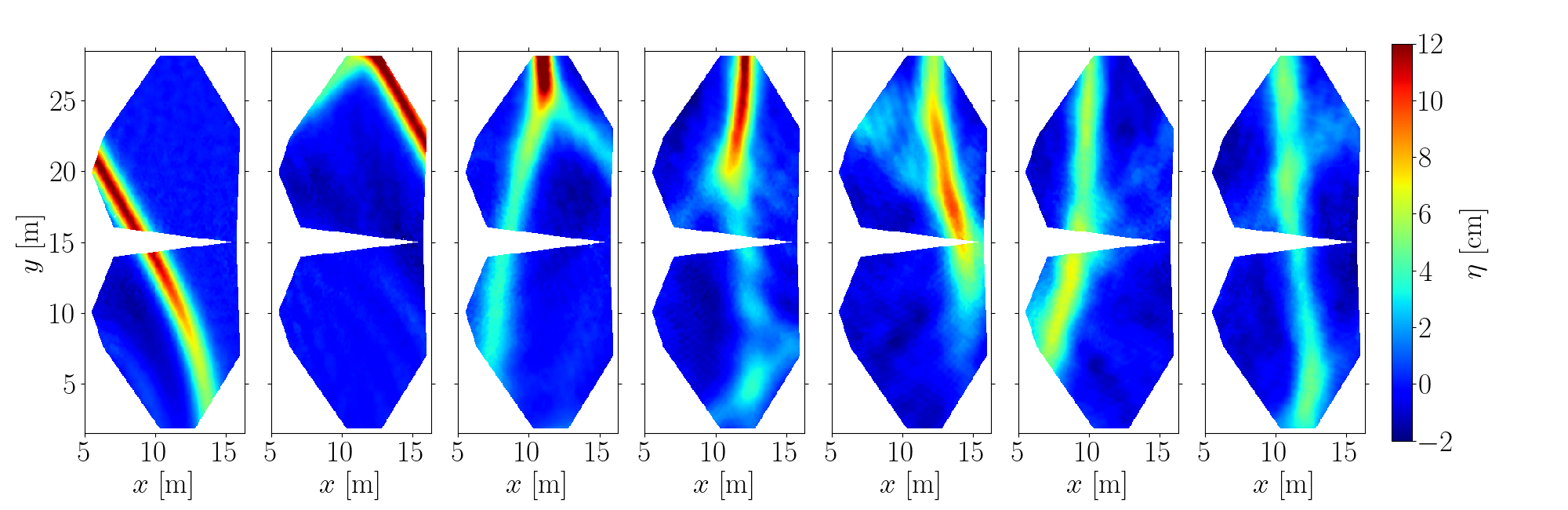}\\
            \vspace*{-16mm}\\
            \hspace*{8mm}
            \textbf{(a)}\hspace{13.1mm}
            \textbf{(b)}\hspace{13.1mm}
            \textbf{(c)}\hspace{13.1mm}
            \textbf{(d)}\hspace{13.1mm}
            \textbf{(e)}\hspace{13.1mm}
            \textbf{(f)}\hspace{13.1mm}
            \textbf{(g)}
            \vspace*{10mm}
            \caption{Stereoscopic measurements of free surface displacements at different times during the propagation of a soliton generated with ${a_0}/{h}=0.35$ and $\theta= 30^\circ$. \textbf{a} $t=7.3$\,s, \textbf{b} $t=11.7$\,s, \textbf{c} $t=24.5$\,s, \textbf{d} $t=35.1$\,s, \textbf{d} $t=49.3$\,s, \textbf{d} $t=60.6$\,s, \textbf{d} $t=77.7$\,s.
            }
            \label{fig:pict_1sol}
        \end{figure}
        
        Fig.~\ref{fig:pict_1sol} shows 7 snapshots of the water surface elevation during the propagation of a single oblique soliton with amplitude $a_0/h=0.35$. As explained in the validation section, we take advantage of a symmetry and the repeatability of experiments to assemble two symmetric experiments and increase the measurement area. In the first experiment, a soliton with $\theta= 30^\circ$ is generated for which the stereoscopic mapping ($y< 15\,$m) is shown in Fig.~\ref{fig:pict_1sol}a. In the second experiment a soliton of same amplitude but with $\theta= - 30^\circ$ is generated. The symmetry of the latter with respect to $y=15\,$m provides the top region of Fig.~\ref{fig:pict_1sol}a. Finally by assembling the two, we obtain the water surface elevation maps over almost the entire width of the tank (from $1.8\,$m to $28.2\,$m).

        \begin{figure}
            \centering
            {\bf (a)}
            \hspace{5cm}
            {\bf (b)}
            \includegraphics[width=0.8\linewidth]{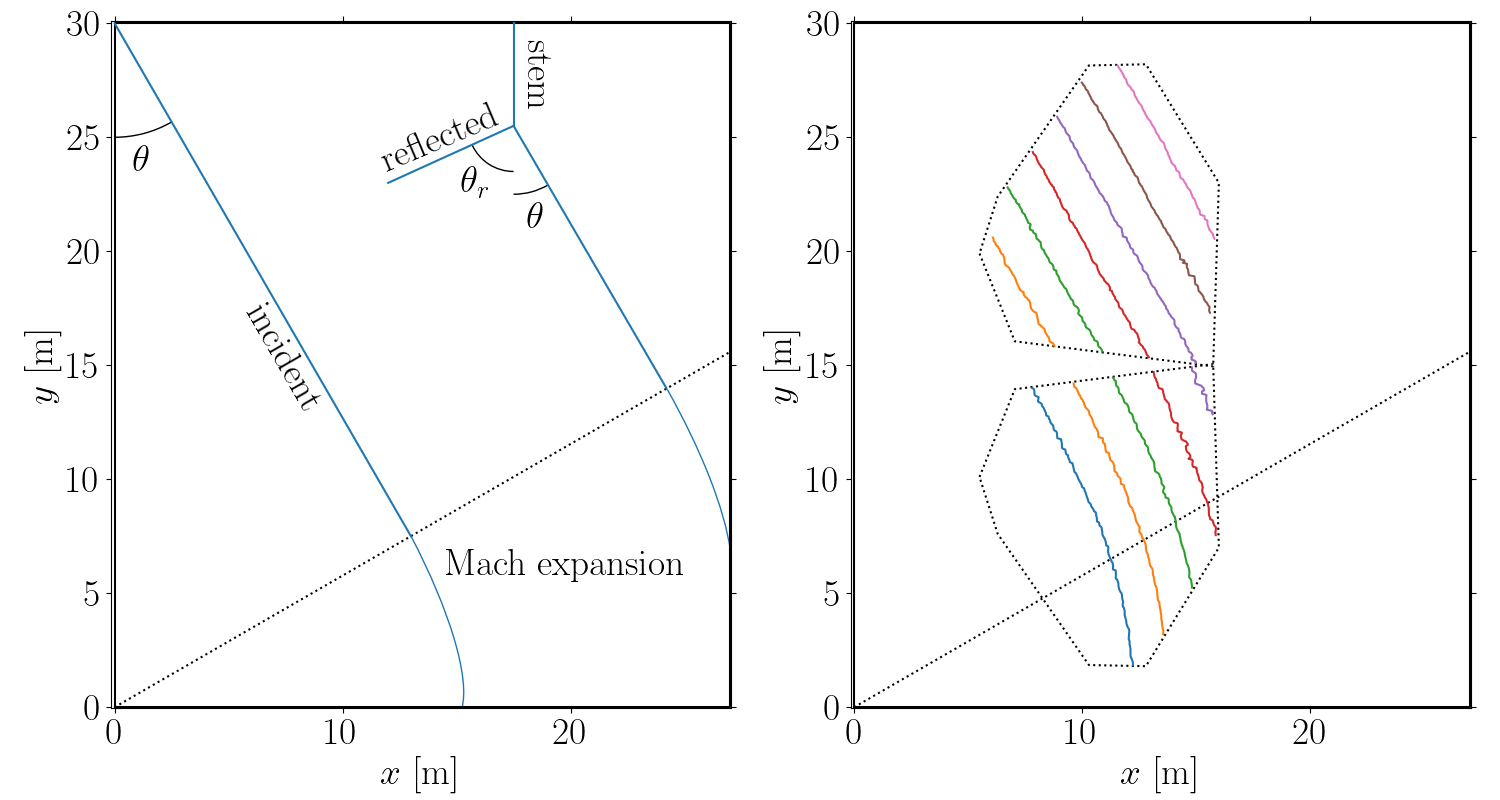}
            \caption{\textbf{a} Schematic diagram of the generation and wall interaction of a soliton (sketch of the crest at two different times). The wave maker generates a soliton of finite width so that a diffracted wave appears next to the $y=0$ wall. The interaction with the $y=30$\,m wall produces a stem and a reflected wave. \textbf{b} Successive positions (with a 0.8\,s time-lapse) of the soliton crest extracted from the stereoscopic measurements (${a_0}/{h}=0.35$, $\theta= 30^\circ$).
            }
            \label{fig:solGen}
        \end{figure}

        At $t=0$ the first piston located at $y=0\,$m generates the first soliton and the last piston located at $y=30\,$m generates its share of the soliton at $t=6.8\,$s. All the other paddles move at intermediate instants according to the law (\ref{eq:sol_gen}). Fig.~\ref{fig:solGen}a illustrates the generation and propagation processes. Due to the finite extension of the wave-maker the crest of an oblique soliton is finite as well and can not extend to the wall located at $y=0$. `Mach expansion' (or diffraction) takes place at this edge of the crest generating a trailing curved crest with decaying amplitude. This Mach expansion is observable in Fig.~\ref{fig:pict_1sol}a at $t=7.3\,$s when the soliton show up in the video measurement region. The soliton crest is straight with constant amplitude from $y\simeq 14\,$m to $21\,$m,  but it is curved and with varying amplitude for $y < 14\,$m. 

        Another process at work is the reflection of the soliton on the $y=30\,$m wall. This interaction, sketched in Fig.~\ref{fig:solGen}a, is defined as `Mach reflection' by \cite{wiegel1964water} and studied in more details by \cite{miles1977obliquely} \eb{and \cite{li2011}}. It generates a \textbf{\sf Y} shape wave which combines the incident wave, a reflected wave (see Fig.~\ref{fig:pict_1sol}b) and a stem (Fig.~\ref{fig:pict_1sol}c). The stem is orthogonal to the wall and it connects the reflected and incident waves. Theoretically the stem amplitude is larger than twice the incidence wave amplitude. The associated motion is unsteady in the frame of reference moving with the stem with the reflected wave building up during propagation. The amplitude and angles of the incident and reflected waves are also in constant evolution. At $t=11.7\,$s, the stem is not visible because it lies out of the field of view of the cameras. However as the wave propagates the lateral extension of the stem grows \citep[e.g.][]{li2011} and during the travel back after the reflection on the end wall (at $x=27\,$m) it can be recorded by the cameras (Fig.~\ref{fig:pict_1sol}c). This type of interaction was analyzed in many studies \citep[]{miles1977resonantly, melville1980,li2011} but never video recorded at such a large scale. Subsequent passages of the main wave are shown in Fig.~\ref{fig:pict_1sol}d-g. The overall trend is to produce a soliton-like wave that propagates in a direction perpendicular to the wave-makers with an amplitude much smaller than the initial wave.

        This illustrates the complex wave dynamics that can emerge in a closed domain from a simple localized line soliton. In other words, an oblique line soliton propagating in a square domain does not remain localized, in contrast with the case of a 1D line soliton (i.e $\theta=0$). For the 30$^\circ$ incidence truncated line soliton shown in Fig.~\ref{fig:pict_1sol}, the growing stems propagating along the side walls tend to cover the whole tank (see Fig.~\ref{fig:pict_1sol}g). promoting a trend for waves to ultimately propagate in the $x$ direction. This will be important for the interpretation of the random cases presented in the two next subsections. 

        \begin{figure}
            \includegraphics[width=.49\linewidth]{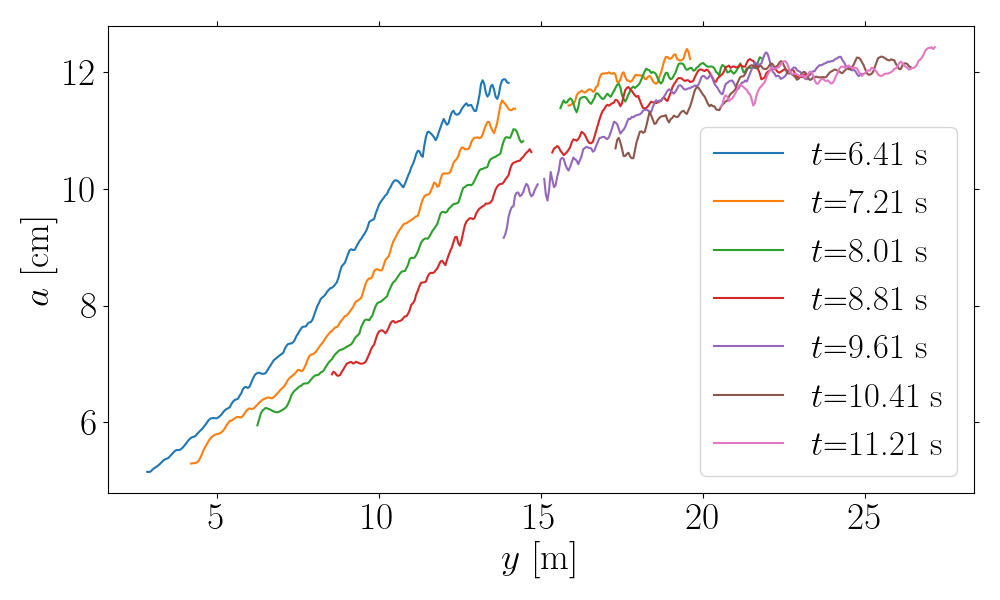}
            \caption{Crest amplitude of the diffracted soliton (${a_0}/{h}=0.35$, $\theta= 30^\circ$) as a function of $y$ at different times. The different curves correspond to the crests shown in Fig.~\ref{fig:solGen}b (same color code as in Fig.~\ref{fig:solGen}).}
            \label{fig:Amp}
        \end{figure}

        \begin{figure}
            \centering
            {\bf (a)}
            \hspace{5cm}
            {\bf (b)}
            \includegraphics[width=\linewidth]{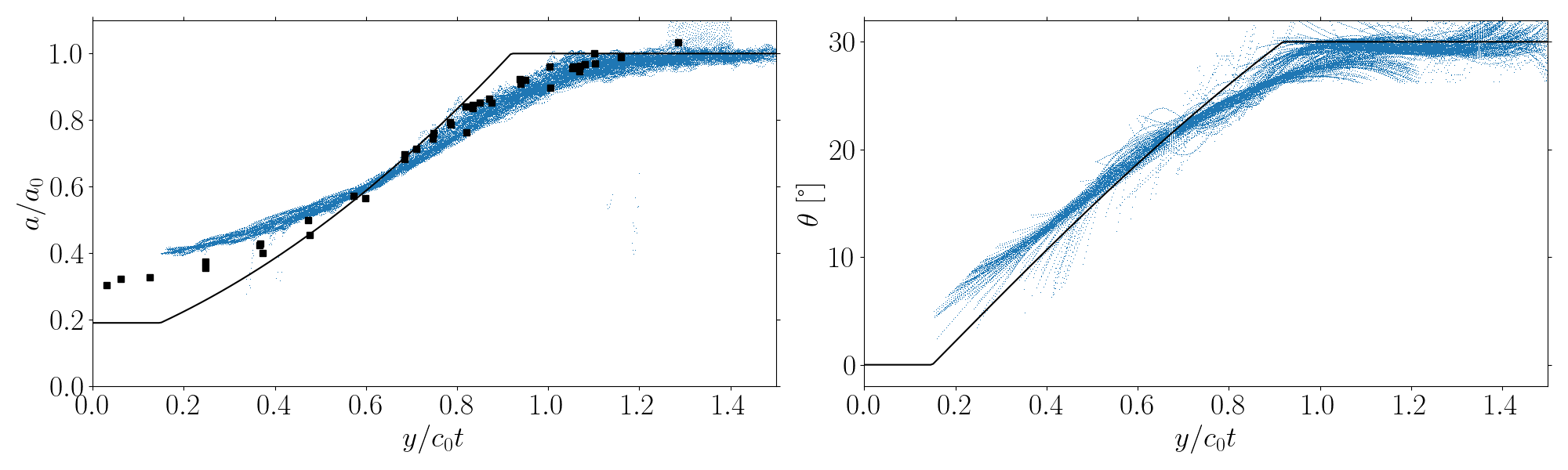}
            \caption{Soliton diffraction, experiment (${a_0}/{h}=0.35$, $\theta=30^\circ$) and KP theory: wave crest \textbf{a} amplitude and \textbf{b} local angle, as a function of ${y}/{t}$, measured by video stereo imaging (blue dots) and by the wave gauges (black squares). Solid lines: analytical predictions from Eqs. (4.17) in \cite{ryskamp2021}.
            }
            \label{fig:Amp_Curve}
        \end{figure}

        \cite{ryskamp2021} studied analytically and numerically the evolution of truncated solitons (with initial finite lateral extension) in the KP framework. The diffraction mechanism described above also takes place in the form of a curved crest with a decreasing amplitude. They show that this diffraction process is self-similar in time: the evolution of the amplitude and curvature only depends on ${y}/{t}$. 
        \TL{Fig.~\ref{fig:solGen}b shows a few successive crest lines of the same wave before the first reflection at the wall occurs. On this one tank travel length the viscous damping can be considered as negligible (see Fig~\ref{fig:sol_keulegan}a).
        These  lines are used to compute the local amplitude and the local angle along the diffracted crest.} 
        Fig.~\ref{fig:Amp} gives the amplitude evolution along the diffracted wave crest as a function of $y$ and time $t$. The gap between curves around the $y=15\,$m location is the gap in the video footprint (see Fig.~\ref{fig:pict_1sol}). At large values of $y$, the crest amplitude is that of the generated soliton. The amplitude along the crest decreases with decreasing $y$. 

        In order to check the self-similar behaviour, the crest amplitude and angle are shown as a function of ${y}/{c_0t}$ in Fig.~\ref{fig:Amp_Curve}. The collapse of all data on the same curve confirms that the crest amplitude evolution is indeed self-similar. The scatter in the video data comes from the uncertainties in crest locations detection. The wave gauge crest amplitude measurements are within the scatter interval of the stereoscopic measurements, confirming that the stereoscopic reconstruction is adapted for such large scale coherent wave motion. The piece-wise theoretical ``expansion wave" prediction of \cite{ryskamp2021} does not exactly fit the data but follows the right trend. The expansion wave solution is based on an approximation of the KP equation postulating slowly evolving wave characteristics in time and space. Obviously this approximation is not valid at short times. Therefore this explains why the measurements of the wave gauges located further away from the video recorded zone are closer to the expansion wave solution. 
        \TL{Note that the theoretical prediction in \cite{ryskamp2021} show a moderate agreement with their numerical results which are qualitatively very similar to our experimental results.}

        The local angle with respect to the $y$ axis along the crest is plotted in Fig.~\ref{fig:Amp_Curve}b. At large $y$ the crest angle is $-30$°, that of the generated soliton. The time evolution is also self-similar  even though the scatter is higher than that of the amplitude. As for the amplitude the theoretical prediction of the local angle is consistent with our observations.

    \subsection{Multi-soliton forcing} \label{MultiSolDyn}

\begin{figure}
            \centering
                 {\bf (a)}
            {\includegraphics[width=0.8\linewidth,viewport=60 680 1250 1250,clip]{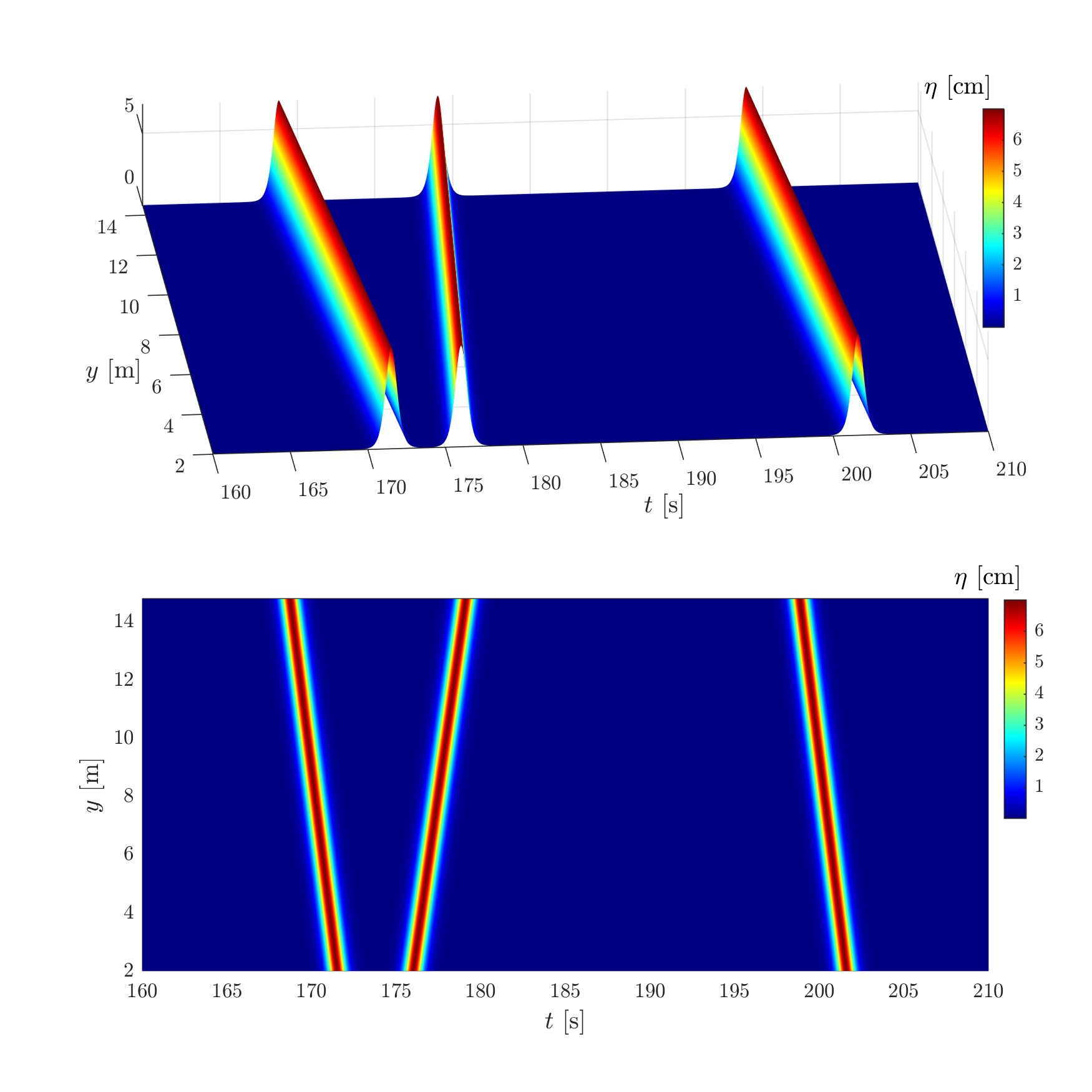}}\\
                 {\bf (b)}
            {\includegraphics[width=0.8\linewidth,viewport=60 680 1250 1250,clip]{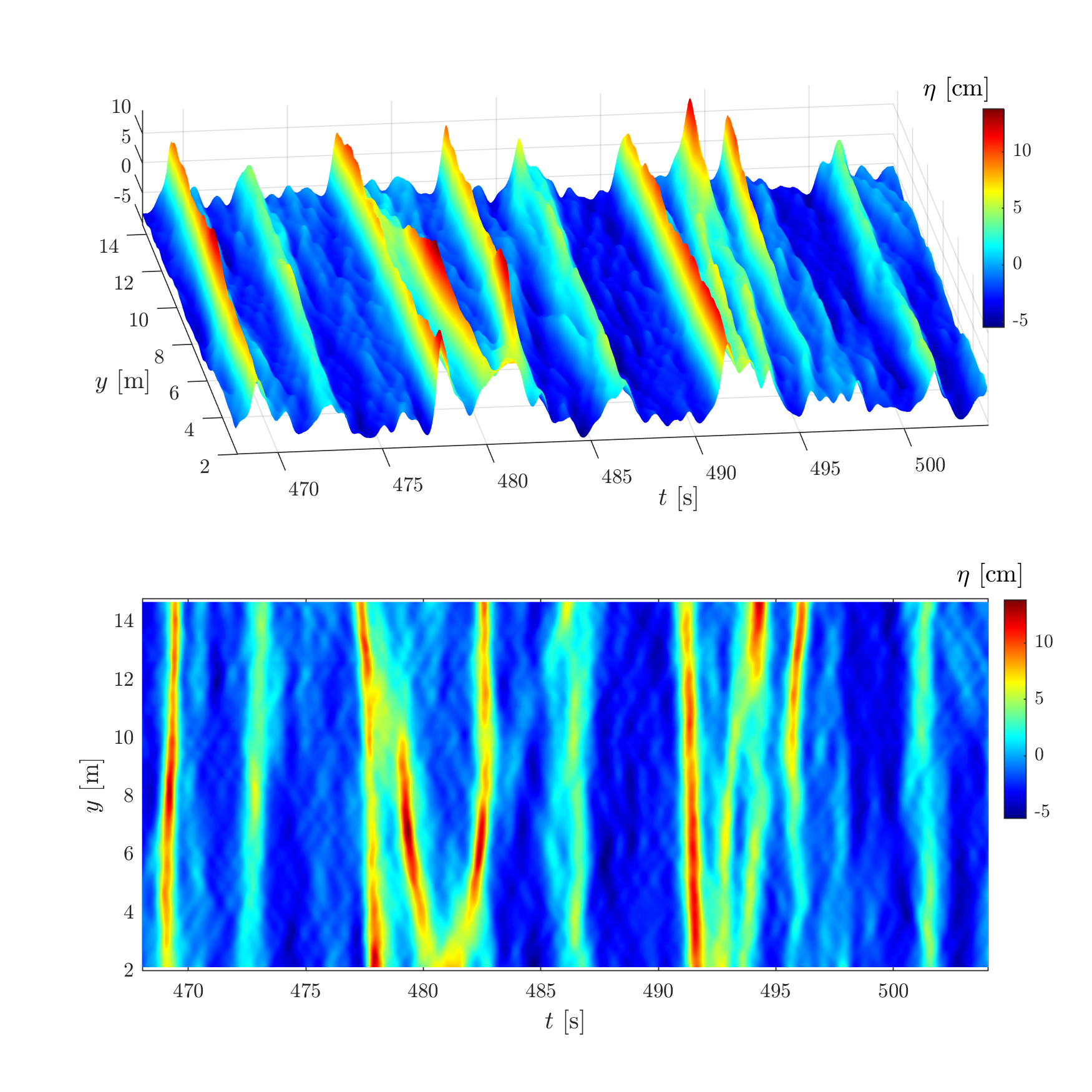}}
            \caption{     
            {\bf a} Example of a short time series of the forcing sequence of the multi-soliton case.
            {\bf b} Portion of the surface elevation field $\eta(t,y)$ at $x=13$\,m measured during the multi-soliton experiment.
            \label{fig:etayt1172}
            }
        \end{figure}

        In this section we analyze an experiment in which solitons are repeatedly generated with a given reduced amplitude ${a_0}/{h}= 0.2$ and with angles chosen randomly with uniform probability between $-30^\circ$ and $30^{\circ}$. Between the forward paddle strokes that generate the solitons, the wave-maker paddles move slowly backwards at constant speed $1.75\,$cm/s in order to \eb{allow} for the next soliton generation. 
        \TL{Short sequences of the forcing wave field and measured wave field are shown in Fig.~\ref{fig:etayt1172}. Due to the technical constraints of the wave-maker excursion, the gas at the generation is very diluted (Fig.~\ref{fig:etayt1172}a).}
        \eb{One soliton is emitted every 21.6 s on average, which corresponds to a 'density' of $1/40$~m$^{-1}$. The generated solitons are then free to interact with the walls and the other waves before they eventually get damped by viscous dissipation. The wave field resulting from this continuous random (in phase and direction, $a_0/h$ = 0.2) soliton forcing is likely denser in solitons (Fig.~\ref{fig:etayt1172}b). Considering that a soliton bounces back on the generation wall typically 10 times before being damped (see Fig.\ref{fig:sol_keulegan}a), the density is $1/4$~m$^{-1}$, typically 10 times higher than the forcing density.}
        \TL{These waves have different amplitudes and speeds, depending on their propagation time since generation, on the multiple interactions with each other, on the interaction with the wave-maker and with the walls.
        }
        
        This random wave field is analysed using a time and space Fourier transform. The multidimensional power spectral density (PSD) $E(k_x,k_y,\omega)$ is defined as: 
        \begin{equation}
            E(k_x,k_y,\omega) = \frac{1}{2\pi L_xL_yT}  \, \left\langle  \left\lvert \iiint \eta(x,y,t) \, \text{Han}(x,y,t) \, e^{i(k_xx+k_yy-\omega t)} \, \text{d}x \text{d}y \text{d}t \right\rvert ^2 \right\rangle  \label{fig:Fourier_Welch}
        \end{equation}
        where $\langle\cdot\rangle$ stands for the ensemble average of PSD estimates on successive time windows of duration $T$. $L_x$ and $L_y$ are the sizes of the measured domain and $\eta$ is the water surface elevation. $\text{Han}$ is a space and time window with Hanning profiles along $x$ or $y$ and that fits the shape of the reconstruction domain. In space domain, $\text{Han}$ is zero at the boundaries of the domain of stereoscopic reconstruction. $\text{Han}$ has also a Hanning profile in time of duration $T$. Such a window is required to avoid spurious effects of the finite extension of the space domain and time interval.

        \begin{figure}
            \centering
            \includegraphics[width=0.3\linewidth,viewport=5 1 250 250,clip]{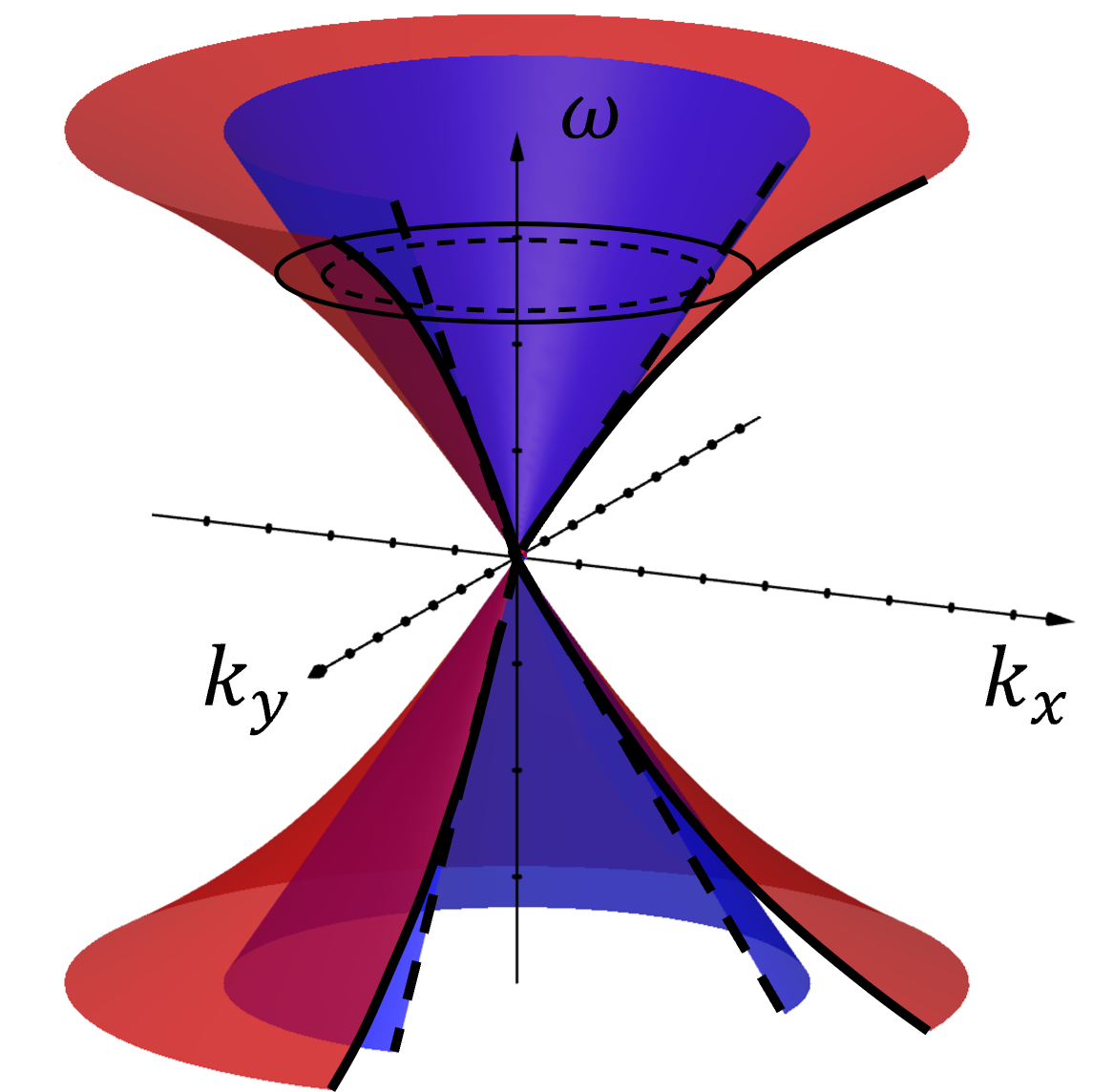}
            \caption{Dispersion relation in the 3D Fourier space. Red surface: linear wave dispersion relation (\ref{lin:disp}). Blue surface: dispersion relation for long waves (\ref{disp:sol0}). For $\omega$ given, in the Fourier space ($k_x,\,k_y$), these dispersion relations are represented by a solid black line for dispersive waves and a dashed black line for long waves.
            }
            \label{fig:Schema-RD}
        \end{figure}
        
        The two dispersion relations for both linear gravity waves and shallow water waves are illustrated in Fig.~\ref{fig:Schema-RD}. The red surface is the dispersion relation for water surface linear gravity waves or Airy relation given by
        \begin{equation}
            \omega = \sqrt{g \, k \tanh{kh}}  \,,
			\label{lin:disp}
        \end{equation}
        where $\omega$ is the angular frequency, $g$ the gravity acceleration, $k$ the norm of the wave number vector $\mathbf k$ of components $(k_x,k_y)$ and $h$ the water depth at rest. The dispersion relation associated to solitons reads
        \begin{equation}
            \omega = c\, k   \,,
			\label{NL:disp}
        \end{equation}
        with $c = \sqrt{gh(1+\frac{a}{h})}$. As a single line soliton propagates in infinite space without altering its shape, all its Fourier modes propagate at the same speed. Thus the dispersion relation of a given soliton is a corresponding line in the $(\mathbf k,\omega)$ space with a slope given by (\ref{NL:disp}). For a collection of independently generated identical line solitons, the power spectrum should be localized on a cone in Fig.~\ref{fig:Schema-RD}. Since solitons propagate at phase speeds $c>c_0$ with $c_0= \sqrt{g \, h}$ their dispersion relation lies inside the cone bounded by
        \begin{equation}
            \omega = c_0 k\,, \text{ with } c_0=\sqrt{gh} \,.
			\label{disp:sol0}
        \end{equation}        
        The dispersion relation (\ref{disp:sol0}) is shown as the blue surface in Fig.~\ref{fig:Schema-RD}. Small amplitude solitons have a signature in the $E(k_x,k_y,\omega)$ space close to the cone (\ref{disp:sol0}) but slightly closer to the $\omega$ axis.

        \begin{figure}
            \includegraphics[width=\linewidth]{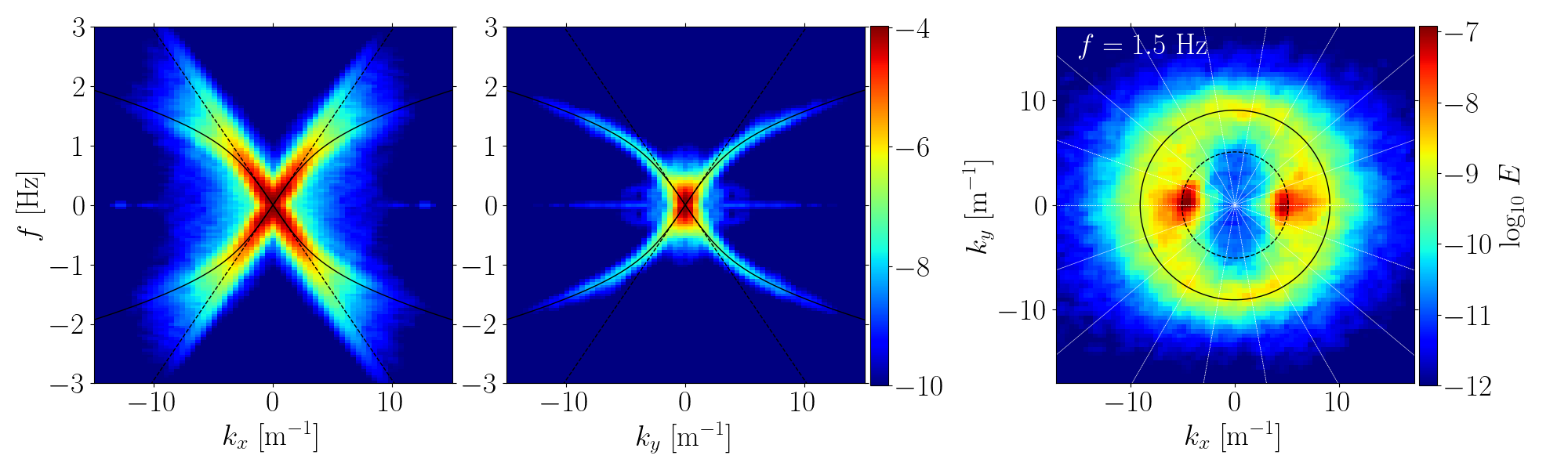}
            \\
            \vspace*{-16mm}\\
            \hspace*{9mm}
            \textcolor{white}{\textbf{(a)}}\hspace{35mm}
            \textcolor{white}{\textbf{(b)}}\hspace{49mm}
            \textcolor{white}{\textbf{(c)}}
            \vspace*{10mm}
            \caption{3D Fourier spectrum $E$ cross-sections at $k_y=0$ (\textbf{a}), $k_x=0$ (\textbf{b}) and $f=1.5$ Hz (\textbf{c}). Multi-soliton experiment (${a_0}/{h}=0.2$ solitons continuously generated, with random incidence between $-30^\circ$ and $30^\circ$). The solid black lines are the traces of the linear wave dispersion relation (\ref{lin:disp}) and the black dashed lines are the traces of the dispersion relation for long waves (\ref{disp:sol0}).
            }
            \label{fig:MultiSol_Sp3D}
        \end{figure}

        In Fig.~\ref{fig:MultiSol_Sp3D}, we show three cross-sections of the $(k_x,k_y,\omega)$ spectrum. The 3D Fourier transform is averaged over $21$ overlapping windows ($50\,$\% overlap) of duration $T = 20$~s starting 9.5~min after the beginning of the experiment. Fig.~\ref{fig:MultiSol_Sp3D}a shows the cross-section taken in the plane $k_y=0$. Only waves that propagate perpendicularly to the wave-maker (in the $x$ direction) leave a signature in this cross-section. The generated solitons travel faster than $c_0$ (dashed black line) evidenced by the high energy straight ridges. Dispersive waves following the Airy relation (\ref{lin:disp}) are also generated with a curved dispersion relation (black continuous line). These dispersive waves arise from the diffraction and reflection of the solitons as described in the previous section and from the backward motion of the paddles. Fig.~\ref{fig:MultiSol_Sp3D}b shows the cross-section taken in the plane $k_x=0$. This provides information on waves that propagate along the wave-makers (in the $y$ direction). The wave energy is distributed along the Airy dispersive relation (black line) and there is no signature of soliton propagation in that direction. 

        Fig.~\ref{fig:MultiSol_Sp3D}c displays a cross-section of the 3D spectra taken at the constant frequency plane $f=1.5$ Hz. This cross-section gives information on the wave propagation incidence. The inner circle and the outer circle are respectively the trace of the dispersion relation for solitons and the dispersion relation for the linear dispersive waves.	Energy of the solitons is located close to the  $k_x$ axis therefore solitons propagation direction is essentially perpendicular to the wave-maker. Even though solitons are generated with directional spreading, the final state is essentially made of non-linear waves propagating in the $x$ direction. As described in section \ref{1SolDyn}, oblique solitons are not stable, they produce stems when interacting with the side walls. The stems build up and tend to force soliton propagation in the $x$ direction. The region $k_x<0$ (respectively $k_x>0$) corresponds to waves that propagate away from the wave-maker (respectively towards the wave-maker). The non-linear waves that propagate away (left red patch on the inner circle) from the wave-maker are more energetic than those propagating towards the wave-maker (right red patch on the inner circle). The reasons are twofold: (i) newly generated solitons propagating away from the wave-maker have experienced less dissipation by bottom friction than those reflected on the $x=27\,$m wall, (ii) the stem generation is associated to the generation of reflected waves that also extract energy to the incident solitons. The uniformly distributed energy on the outer circle of this cross-section shows that dispersive waves are isotropic in contrast with the solitons.

    \subsection{Random forcing} \label{RandDyn}
	       
          \begin{figure*}
            \includegraphics[width=\linewidth]{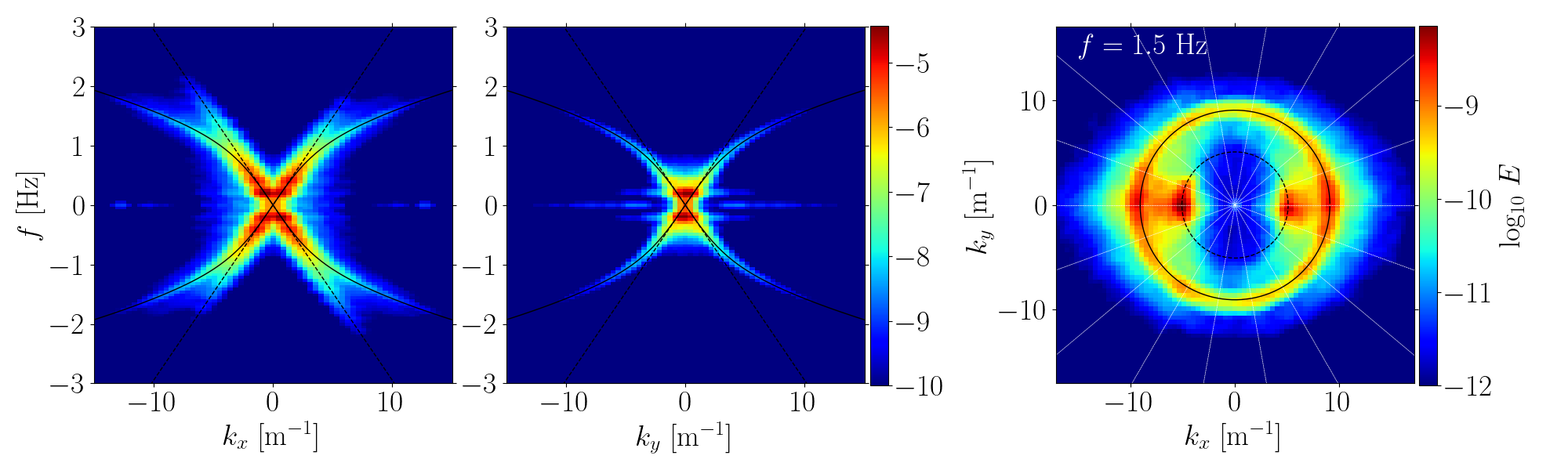}
            \\
            \vspace*{-16mm}\\
            \hspace*{9mm}
            \textcolor{white}{\textbf{(a)}}\hspace{35mm}
            \textcolor{white}{\textbf{(b)}}\hspace{49mm}
            \textcolor{white}{\textbf{(c)}}
            \vspace*{10mm}
            \caption{3D Fourier spectrum $E$ cross-sections at $k_y=0$ (\textbf{a}), $k_x=0$ (\textbf{b}), $f=1.5$ Hz (\textbf{c}). {\sc jonswap} experiment B. The solid black lines are the traces of the linear wave dispersion relation (\ref{lin:disp}), dashed black lines are traces of the dispersion relation for long waves (\ref{disp:sol0}).
            }
            \label{fig:3D_spectra}
        \end{figure*}
        
        As mentioned in section~\ref{sect:generation:random}, a {\sc jonswap} forcing is a convenient way to produce a wave random state in the tank. For {\sc jonswap} experiment B, a 14 min recording is used, starting 8~min after the beginning of the wave generation. We apply space and time Fourier transforms (\ref{fig:Fourier_Welch}) on $81$ overlapping windows (with $50\,$\% overlap) of duration 20\,s. Three cross-sections of the full spectra are shown in Fig.~\ref{fig:3D_spectra}. The first cross-section (Fig.~\ref{fig:3D_spectra}a) is the plane $k_y=0$ to detect waves propagating in the $x$ direction (wave crests parallel to the wave-maker). This cross-section is somewhat similar to the previous case of direct forcing of solitons. Solitons emerge spontaneously as can be seen from the concentration of energy on the straight line, although the forcing was not designed to specifically generate solitons. A large part of the waves are linear dispersive waves following the Airy dispersion relation (\ref{lin:disp}). The second cross-section of the spectrum shown in Fig.~\ref{fig:3D_spectra}b is taken at $k_x=0$ to detect waves progagating in the $y$ direction. In this direction of propagation, only dispersive waves are evidenced as for the previous case of multi-soliton generation. The last cross-section, shown in Fig.~\ref{fig:3D_spectra}c is the plane $f=$ 1.5 Hz. This plot is consistent with the two previous cross-sections and evidence the presence of solitons propagating mainly in the $x$ direction. Dispersive waves propagate in all directions, still presenting some anisotropy resulting from the $\pm 45^\circ$ directivity of the forcing.

        The global picture in the case of random forcing is rather similar to that of multiple soliton forcing. The wave field is made of a mixture of solitons and dispersive waves in both cases. The solitons propagate with small directionnal spreading around the $x$ direction because of the instability of oblique solitons, promoted by Mach reflection in such a finite domain. Dispersive waves are much more isotropic.

\section{Conclusions}

    Water surface wave motions mapping  (maximum vertical elevation $\sim 25$\,cm, with a typical 5\,mm accuracy) was achieved in a ($27$\,m$\times 30$\,m$\times 0.35$\,m) wave tank over about 100\,m$^2$ by stereo-video, with a time resolution of $25\,$Hz, over times of several tens of minutes. 
    \TL{The stereo-video system is based on the cross-correlation of images from two cameras for which  patterns produced by floating particles are necessary.}
    It was observed that wave breaking strongly redistributes the particles into dense patches and leaving holes in the surface coverage. This precludes accurate measurements in wave breaking situations. 
     
    For the propagation of a single oblique ($\theta=30^\circ$) line soliton Mach reflection and Mach expansion are evidenced. On the one hand, Mach reflection produces a stem wave perpendicular to the side wall facing the soliton. On the other hand the crest of the diffracted wave resulting from Mach expansion tends to attach perpendicularly to the side walls as theoretically predicted. The measured wave amplitude decay and local crest angles for the Mach expansion confirm the theoretical predictions of \cite{ryskamp2021}. \eb{A close inspection of soliton interaction, including stem stability, that might challenge the validity of the assumptions underlying the K-P equations, is still under progress.}
    
    For the first time to our best knowledge, random 2D soliton gas is generated in a laboratory wave tank. This is illustrated with two different cases. One gas is forced by repeatedly producing line solitons with random incidence ($-30^\circ<\theta<30^\circ$). Both Mach reflections and Mach expansions take place and the solitons tend to propagate in a direction perpendicular to the wave-makers (wave crests perpendicular to the side walls). The other gas is produced by random waves resulting from wave-maker motions complying with a {\sc {\sc jonswap}} spectrum with random incidence ($-45^\circ<\theta<45^\circ$). Waves travelling faster than Airy waves are identified as solitons, naturally emerging from interactions between wave frequency components and with the walls. \eb{This emergence of soliton gas is also attested by the highly skewed {\sc pdf} of the surface elevation which spontaneously emerges from the Gaussian forcing.} In this case also, most solitons propagate preferentially perpendicularly to the wave-makers. 

    \TL{
    Further investigations are required in order to estimate the amount of solitons (i.e. density) in the various gases. 
    Since the spectral signature of solitons overlaps that of dispersive waves, the Fourier analysis is probably not the relevant tool for that purpose. The most efficient tool for soliton quantification in mono-directional propagation is the  Scattering Transform \cite[e.g.][]{redor2021}. Extension of this method to 2D flows is still to be done.
    }

\section*{Acknowledgements} 

    TL acknowledges the support of the french Ministry of Higher Education and Research through the PhD grant.
    The authors are grateful to ARTELIA company for making their wave tank available and fully operational. Our many thanks to M. Lagauz\`ere for her contribution to the wave-maker control program and to the synchronization of the data acquisition.

\section*{Declarations}

\subsection*{Ethical Approval}
not applicable

\subsection*{Funding }
This study has received funding from Agence Nationale de la Recherche project SOGOOD (ANR-21-CE30-0061) and from the Simons Foundation through the Simons Collaboration on Wave Turbulence.

\subsection*{Availability of data and materials  }
Data available upon request. Video rectification was achieved with the UVMAT software that can be downloaded at:
\noindent http://servforge.legi.grenoble-inp.fr/projects/soft-uvmat.

\section*{Appendix} \label{Appendix}
\begin{figure}
        \centering
        \includegraphics[width=0.49\linewidth]{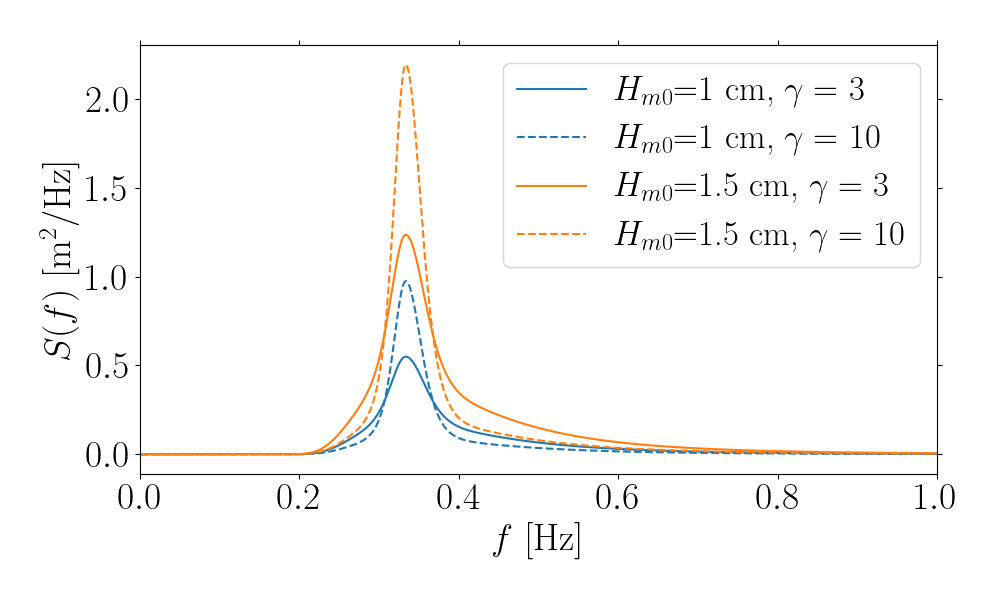}
        \includegraphics[width=0.49\linewidth]{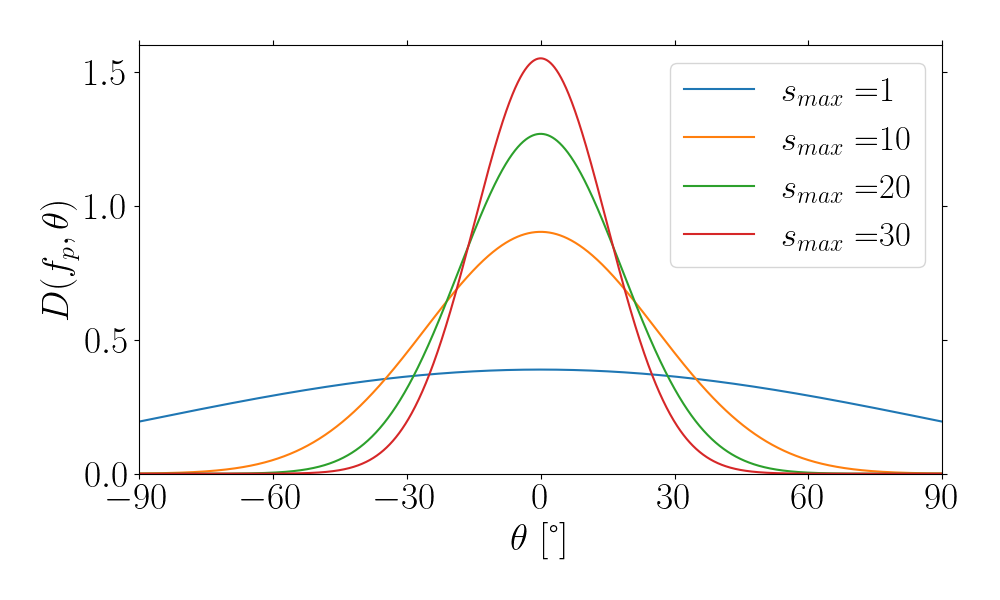}
        \caption{\textbf{a} Fourier power density spectra  (\ref{eq:JONSWAP_S(f)})
        with the same peak frequency $f_p=0.33$~Hz and various $H_{m0}$ and $\gamma$ given in the insert.
        \textbf{b} Angular distribution at $f_p$ for four values of $s_{\max}$ given in the insert, to which correspond  different values of the angle standard deviation (\ref{eq:StandDev_D(f,theta)}): $\sigma_\theta=14.7^\circ$ (red), $\sigma_\theta=17.9^\circ$ (green), $\sigma_\theta=25^\circ$ (orange), $\sigma_\theta=47.4^\circ$ (blue).}
        \label{fig:JONSWAP}
    \end{figure}

\eb{\cite{hasselmann1973measurements}, through the analysis of the data collected during the JOint North Sea WAve Observation Project ({\sc jonswap}), found a typical wave spectrum that can match approximately many sea states. 
It reads:
    }
    \begin{equation}
        S(f)= \alpha H_{m0}^2 \frac{f_p^4}{f^5} e^{-\frac{5}{4}\left(  \frac{f_p}{f} \right)^4} \gamma^\beta
        \label{eq:JONSWAP_S(f)}
    \end{equation}
 \TL{
   with
   }
    \begin{equation}
        \beta=\exp{\left( -\frac{(f-f_p)^2}{2\sigma^2 f_p^2}\right)}
        , 
        \qquad
        \alpha=\frac{0.064}{0.23+0.0336\gamma-\frac{0.185}{1.9+\gamma}} \,,
    \end{equation}
    \begin{equation}
        \sigma=\left\{ 
            \begin{matrix} 
            0.07\quad \mathrm{for} \quad f\leq f_p \\ 0.09\quad \mathrm{for} \quad  f > f_p \end{matrix} 
            \right. \,,
    \end{equation}
\TL{
    where $H_{m0}$ is a significant wave height equal to four times the standard deviation of the surface elevation, $f_p$ is the peak frequency and $\gamma$ is the peak enhancement factor. Examples are shown in Fig.~\ref{fig:JONSWAP}{\bf a}. 
    }
\eb{ Nowadays, the {\sc jonswap} spectrum is a common tool for modeling the sea state in deep water. Here we use it as a standard random wave generation procedure.
}

\TL{
    The angular spreading distribution is 
    \cite[e.g][]{goda1999comparative}
}    
    \begin{equation}
        D(f,\theta)=\frac{\text{\textGamma}(s+1)}{2\sqrt{\pi}\text{\textGamma}(s+1/2)}\cos^{2s}\left(\frac{\theta-\theta_0}{2}\right)
        \label{eq:JONSWAP_D(f,theta)}
    \end{equation}
    with
    \begin{equation}
        s=\left\{ 
        \begin{matrix} 
            s_{\max}(f/f_p)^5\quad \mathrm{for} \quad f\leq f_p \\ 
            s_{\max}(f/f_p)^{-5/2}\quad \mathrm{for} \quad f > f_p \end{matrix} 
            \right. \,,
    \end{equation}
    \begin{equation}
        \sigma_\theta ^2 = \int_{-\pi/2}^{\pi/2} \theta^2 D(f_{p}, \theta)  \mathrm{d}\theta \,,
        \label{eq:StandDev_D(f,theta)}
    \end{equation}
    \TL{
    where $\theta_0$ is the principal angle and the angular range is set by $s_{\max}$. Examples are shown in 
Fig.~\ref{eq:JONSWAP_D(f,theta)}b. In our case, the angular spreading distribution is truncated at $\theta=\pm45^\circ$.
}
    
\eb{
The resulting spectrum $E(f,\theta)=S(f)D(f,\theta)$ is discretized into $10^5$  components with random phases. The surface elevation at the wave-maker is computed through the inverse Fourier transform, setting linear wave components   
 with amplitudes $a_i$.
The paddle stroke $X_i$ of each component 
is computed as
\cite[][]{hughes1993physical}
\begin{equation}
    X_i = a_i \frac{\sinh(2 k_i h)+2 k_i h}{4\sinh^2(k_i h)} 
\end{equation}
where each wave number  ($k_i$) fulfils the dispersion relation
(\ref{lin:disp}).
}

\bibliographystyle{spbasic}
\bibliography{Biblio/references.bib}

\end{document}